\documentclass[aps,prd,superscriptaddress,showkeys,showpacs,nofootinbib,10pt,twocolumn,longbibliography,floatfix]{revtex4-2}

\usepackage[english]{babel}

\usepackage{color}
\usepackage{graphicx}
\usepackage{makecell}
\usepackage{booktabs}

\usepackage{listings}

\usepackage{amsmath,amssymb,empheq}
\usepackage{appendix}

\allowdisplaybreaks

\usepackage[breaklinks=true]{hyperref}
\hypersetup{
	colorlinks=true,
	linkcolor=blue,
	filecolor=magenta,      
	urlcolor=blue,
	citecolor=red,
}

\usepackage{orcidlink}

\usepackage{mathrsfs}
\usepackage[normalem]{ulem}

\usepackage{placeins}
\usepackage[utf8]{inputenc}

\input epsf
\usepackage{mathtools}
\usepackage{amsfonts}
\usepackage{upgreek}
\usepackage{latexsym}

\usepackage{afterpage}
\usepackage{bm}

\usepackage{siunitx}
\usepackage{multirow}

\usepackage{stackengine}
\usepackage{scalerel}
\usepackage{xcolor}

\begin{document}
{\hfill MS-TP-25-15}
\title{Reheating ACTs on Starobinsky and Higgs inflation}

\author{D.S.~Zharov\,\orcidlink{0009-0000-6439-0958}}
\email{dmitriy.zharov.02@knu.ua}
\affiliation{Physics Faculty, \href{https://ror.org/02aaqv166}{Taras Shevchenko National University of Kyiv}, 64/13, Volodymyrska Street, 01601 Kyiv, Ukraine}

\author{O.O.~Sobol\,\orcidlink{0000-0002-6300-3079}}
\email{oleksandr.sobol@knu.ua}
\affiliation{Institute for Theoretical Physics, \href{https://ror.org/00pd74e08}{University of M\"{u}nster}, Wilhelm-Klemm-Stra{\ss}e 9, 48149 M\"{u}nster, Germany}
\affiliation{Physics Faculty, \href{https://ror.org/02aaqv166}{Taras Shevchenko National University of Kyiv}, 64/13, Volodymyrska Street, 01601 Kyiv, Ukraine}

\author{S.I.~Vilchinskii\,\orcidlink{0000-0002-9294-9939}}
\affiliation{D\'{e}partement de Physique Th\'{e}orique and Center for Astroparticle Physics, \href{https://ror.org/01swzsf04}{Universit\'{e} de Gen\`{e}ve},  24 quai Ernest Ansermet, 1211 Gen\`{e}ve 4, Switzerland}
\affiliation{Physics Faculty, \href{https://ror.org/02aaqv166}{Taras Shevchenko National University of Kyiv}, 64/13, Volodymyrska Street, 01601 Kyiv, Ukraine}	

\date{\today}

\begin{abstract}
In the recent sixth data release (DR6) of the Atacama Cosmology Telescope (ACT) collaboration, the value of $n_{\rm s}=0.9743 \pm 0.0034$ for the scalar spectral index is reported, which excludes the Starobinsky and Higgs inflationary models at $2\sigma$ level. In this paper, we perform a Bayesian inference of the parameters of the Starobinsky or Higgs inflationary model with non-instantaneous reheating using the Markov chain Monte Carlo method. For the analysis, we use observational data on the cosmic microwave background collected by the Planck and ACT collaborations and on baryonic acoustic oscillations from the Dark Energy Spectroscopic Instrument (DESI) collaboration. The reheating stage is modelled by a single parameter $R_{\text{reh}}$, which contains a combination of the reheating temperature $T_{\text{reh}}$ and the effective equation of state of matter during reheating $\bar\omega_{\text{reh}}$. Using the modified Boltzmann code \texttt{CLASS} and the \texttt{cobaya} software with the \texttt{GetDist} package, we perform a direct inference of the model parameter space and obtain their posterior distributions. Using the Kullback--Leibler divergence, we estimate the information gain obtained from the observed data: In the proposed parameterization, we get $8.66$ bits of information about the amplitude of the inflaton potential and $2.52$ bits of information about the reheating parameter.  
Inclusion of the ACT DR6 data provides $75\%$ more information about the reheating stage compared to analysis without ACT data. In addition, we draw constraints on the reheating temperature and the average equation of state. While the former can vary within $10$ orders of magnitude, values in the $95\,\%$ credible interval indicate a sufficiently low reheating temperature; for the latter there is a clear preference for values greater than $0.5$, which means that the conventional equations of state for dust $\omega=0$ and relativistic matter $\omega=1/3$ are excluded with more than $2\sigma$ level of significance. Nevertheless, there still is a big part of parameter space where Starobinsky and Higgs inflationary models exhibit a high degree of consistency with the latest observational data, particularly from ACT DR6. Therefore, it is premature to reject these models.
\end{abstract}

\maketitle

%%%%%%%%%%%%%%%%%%%%%%%%%%%%%%%%%%%%%%%%%%%%%%%%%%%%%%%%%
%%%%%%%%%%%%%%%%%%%%%%%%%%%%%%%%%%%%%%%%%%%%%%%%%%%%%%%%%
\section{Introduction}
\label{sec:introduction}
%%%%%%%%%%%%%%%%%%%%%%%%%%%%%%%%%%%%%%%%%%%%%%%%%%%%%%%%%
%%%%%%%%%%%%%%%%%%%%%%%%%%%%%%%%%%%%%%%%%%%%%%%%%%%%%%%%%

Constructing a coherent model of the Universe has long been one of humanity's central scientific goals. In the 1990s, the $\Lambda$CDM model emerged as the standard cosmological framework, describing a spatially flat Universe undergoing the Big Bang expansion~\cite{Weinberg:2008zzc,dodelsonModernCosmology2021}. Predictions of the $\Lambda$CDM model show a high degree of agreement with observational data, in particular, the anisotropy and polarization of the cosmic microwave background (CMB), baryon acoustic oscillations (BAO), 
large-scale structure (LSS) etc. Each measurement imposes constraints on specific model parameters that describe the dynamics of the Universe evolution, such as the Hubble parameter $H_0$, the density parameters $\Omega_i$ for different components, and parameters of primordial perturbations. 
%Using these measurements, the best values of the model parameters are determined by means of the joint Bayesian inference.

Undoubtedly, one of the most important sources of information about the evolution of our Universe is the CMB, the spectrum of which was first measured by the COBE satellite~\cite{smoot1992structure-837}, then refined by the WMAP satellite~\cite{spergel2003firstyear-505}, and the latest data on which were collected by the Planck Collaboration~\cite{collaboration2020planck-5b8, collaboration2020planck-a3c, collaboration2019planck-584}, BICEP/Keck Collaborations~\cite{collaboration2021improved-f44}, and the Atacama Cosmology Telescope (ACT)~\cite{louis2025atacama-980}. 
%The measurements of the latter two give potentially contradictory results with DESI BAO measurements.
The joint analysis of the parameter space of the $\Lambda$CDM model using CMB observations by Plank, BICEP/Keck, and ACT collaborations and BAO data from the second data release (DR2) of DESI collaboration~\cite{collaboration2025desi-e55} imposes unexpected constraints on the spectral index of primordial scalar perturbations spectrum $n_\text{s}=0.9743 \pm 0.0034$ \cite{calabrese2025atacama-a33} for the CMB pivot scale of $k_\ast = 0.05 \, \text{Mpc}^{-1}$. This new value is significantly higher from the value $n_s = 0.965 \pm 0.004$ reported in Planck 2018 data release~\cite{Planck:2018vyg}. Reference~\cite{collaboration2025desi-act} shows that the DESI DR2 BAO data exhibit a $1.4\sigma\text{--}3.2\sigma$ discrepancy with the Planck and ACT results which still has to be understood.\footnote{Note that results of different observations are sometimes in contradiction between each other; e.g., the renowned $H_0$ and $\sigma_8$ tensions~\cite{leizerovich2023tensions-540}. Joint analysis of mutually inconsistent datasets has to be performed with caution~\cite{steinhardt2025dark-b8f}.} 

Talking about initial conditions required by the $\Lambda$CDM model, they appear to be rather unnatural (or unlikely) which is usually formulated as puzzles of the hot Big Bang model: the horizon problem, the flatness problem, the monopoles problem, and the problem of the origin of initial inhomogeneities (for more details see, e.g., textbooks \cite{Weinberg:2008zzc,dodelsonModernCosmology2021}). 
To solve these problems, a paradigm of inflation was proposed, the main idea of which is the presence of accelerated expansion of the Universe which occurs before the radiation dominated era \cite{starobinsky1980new-7c9,guth1980inflationary-4c7,Linde:1981mu,Starobinsky:1982ee,Albrecht:1982wi,Linde:1983gd}. At this stage, the universe is dominated by the inflaton scalar field whose effective equation of state is close to that of the vacuum, $p \approx - \rho$; therefore, the universe expands rapidly and eventually becomes homogeneous on large scales and spatially flat with high accuracy. Notably, inflation is also able to produce initial inhomogeneities through the amplification of vacuum fluctuations of the inflaton and metric~\cite{Starobinsky:1979ty,Mukhanov:1981xt,Mukhanov:1982nu,Guth:1982ec,Hawking:1982cz,Bardeen:1983qw}. Spectral properties of these perturbations depend on the shape of the inflaton potential.

One of the most important drawbacks of inflationary paradigm is the potentially infinite number of models which, being appropriately fine-tuned, can describe any observable data. A large but still incomplete list of inflationary models proposed in the literature is given in Ref.~\cite{martin2013encyclopaedia-a01}. Although most of them require an introduction of a new scalar field which is not present in the Standard Model, there are a few well-motivated and economical models which do not require that. One such model was proposed by A.~Starobinsky~\cite{starobinsky1980new-7c9} and represents the simplest $f(R)$ modification of the gravitational sector where the usual Einstein--Hilbert term is supplemented by the $R^2$ term. Here, the scalar degree of freedom playing the role of the inflaton arises from gravity. Another model is the Higgs inflation where the Standard-Model Higgs field is nonminimally coupled to gravity~\cite{bezrukov2008standard-957}. Despite the different physical nature of these models both imply in the Einstein frame the similar inflaton effective potential and lead to the same predictions for the spectral index $n_s$ of primordial scalar perturbations and to a tensor-to-scalar ratio $r$, which, until recently, were always favored by observations~\cite{collaboration2020planck-a3c,martin2013encyclopaedia-a01,martin2024cosmic-687,Chakraborty:2023ocr}.
More recent data from the ACT suggest a higher value for $n_s$ and find that Starobinsky and Higgs inflation models are on the $2\sigma$ bound of the combined constraints if the number of $e$-folds before the end of inflation when the CMB pivot scale crosses the horizon is around $N_\ast=60$. However, $N_\ast$ is not an independent model parameter. It depends on both inflationary model and postinflationary universe expansion.  

Indeed, to build a complete picture of the evolution of the early universe, it is also very important to understand how the inflationary stage transitions into the radiation dominated era. At the end of inflation, it is generally believed that the inflaton oscillates around the minimum of its potential, gradually decaying and transferring energy to the relativistic plasma. This post-inflationary process, which fills the universe with ordinary matter, is known as the reheating. Traditional approaches to reheating typically involve the production of Standard Model particles via perturbative decay of the inflaton field and various types of parametric resonances~\cite{Kofman:1997yn,Allahverdi:2010xz,Amin:2014eta}. Through mutual interactions, the components of the early Universe reach a state of thermodynamic equilibrium, marking the onset of the standard Big Bang cosmology~\cite{Weinberg:2008zzc}.

Usually, the literature neglects to consider the post-inflationary reheating stage, considering it instantaneous. However, as shown by the authors of~\cite{martin2024cosmic-687}, the observational data on CMB and BAO already carry a certain amount of information about this stage ($1.3\pm0.18$ bits). This means that interpreting the observational data without taking into account the reheating can lead to a distortion of the picture of the universe evolution.

% However, in order to implement the joint analysis, it is necessary to reduce to a common basis the observational data of different missions that use different methodologies and are determined by different parameters. For this purpose, it is necessary to assume a certain cosmological model. It may turn out that within this model, measurements made by different methodologies will be contradictory. In this case, the question arises as to how to interpret these contradictions. The authors of Ref.~\cite{steinhardt2025dark-b8f} propose three possible options: the error of one or more methodologies, significant underestimation of systematic errors, and the use of an inaccurate model. 

% The current most accurate observational data are mutually contradictory with respect to some cosmological parameters within the $\Lambda$CDM model, in particular those related to the Hubble parameter $H_0$, the spatial curvature $\Omega_\text{K}$ and the amplitude of the matter spectrum $\sigma_8$ \cite{leizerovich2023tensions-540}, and in the light of the recent second release (DR2) of the BAO data of the Dark Energy Spectroscopic Instrument collaboration (DESI) \cite{collaboration2025desi-e55}, the dark energy state of matter parameter $\omega_\text{DE}$. These contradictions play a serious role in motivating researchers to seek to extend the physics of the standard cosmological model.

In order to get physical intuition about the newest ACT constraints, it is instructive reformulate them in terms of physical parameters. In this paper, we consider the $\Lambda$CDM model with Starobinsky or Higgs inflation followed by non-instantaneous reheating, perform a Bayesian analysis of its parameter space, and, within the framework of this model, propose a different interpretation of the reported value of the spectral index, namely, that this value of $n_\text{s}$ provides us with information about the reheating stage. Our main conclusion is that taking into account non-instantaneous reheating, Starobinsky and Higgs inflation models are still viable candidates for a realistic inflationary model.

This paper is organized as follows. In Section~\ref{sec:model} we give a brief theoretical background on Starobinsky and Higgs inflationary models and the reheating stage. In Section~\ref{sec:methods} we formulate the mathematical problem and outline the features of the numerical methods used. In Section~\ref{sec:results} we present the numerical results on Bayesian parameter inference and analyze them. Section~\ref{sec:conclusion} is devoted to conclusions. Throughout, we will work in the natural system of units $c=\hbar=k_B=1$ and use the Planck mass $M_{\text{Pl}}=1.22\times 10^{19}$~GeV as a basis energy unit.

\section{Model}
\label{sec:model}

\subsection{Starobinsky and Higgs inflationary models}

Among hundreds of inflationary models (see, e.g., Ref.~\cite{martin2013encyclopaedia-a01} for a review), the Starobinsky and Higgs models occupy special places. 

The Starobinsky model~\cite{starobinsky1980new-7c9} was one of the first models of inflation to be proposed. Over the last decades, this model has shown a very high degree of consistency with data, and with each new data release, it has been increasingly confirmed. It arises from natural physical considerations of the generalization of the Einstein--Hilbert action. It is characterized by an action that contains the Einstein-Hilbert term together with an additional term that depends on the second order of scalar curvature
\begin{equation}
S = \int d^4x \sqrt{-g} \left[ -\frac{M_{\text{Pl}}^2}{16\pi} R + \frac{R^2}{6M^2} \right]\, ,  
\end{equation}
where \( R \) denotes the scalar Ricci curvature, \( g \) is the determinant of the metric \(g_{\mu\nu}\), and \( M \) is the free parameter of the Starobinsky model expressed in units of mass. 

Switching to the Einstein frame by means of the conformal transformation, this action can be rewritten in a form that explicitly includes a scalar field
\begin{equation}
S = \int d^4x \sqrt{-\bar{g}} \left[- \frac{M_{\text{Pl}}^2}{16\pi} \bar{R} + \frac{1}{2} \partial^\mu \phi \partial_\mu \phi - V(\phi) \right]\,,
\end{equation}
where the metric and Ricci scalar with bars denote the corresponding quantities in the Einstein frame (we will omit bars in what follows) and \( V(\phi) \) is the potential of a scalar field of the form
\begin{equation}
    V(\phi) = V_0 \left[ 1 - \exp\bigg( - 4\sqrt{\frac{\pi}{3}} \frac{\phi}{M_{\text{Pl}}} \bigg) \right]^2\, 
    \label{eq:potential}
\end{equation}
with \( V_0 = \frac{3}{32\pi} M_{\text{Pl}}^2 M^2\). 

The Higgs inflationary model~\cite{bezrukov2008standard-957} considers the Standard Model Higgs field $h$ to be the inflaton. It is nonminimally coupled to gravity through the $\propto h^2 R$ term. The corresponding action has the form:
\begin{equation}
	\label{a1}
		S\!=\!\!\int\!\! d^4x \sqrt{-g}\Big[\!-\!\frac{M_{\text{Pl}}^2}{16\pi}\Big(1+\frac{8\pi \xi_h h^2}{M_\text{Pl}^2}\Big)R\!+\!\frac{1}{2}\partial_\mu h\partial^\mu h\!-\!\frac{\lambda}{4}h^4\Big],
\end{equation}
where $\lambda$ is the quartic self-coupling constant and $\xi_h$ is the nonminimal coupling constant. Switching again to the Einstein frame, one gets for the canonically normalized inflaton field exactly the same potential as the one in Eq.~\eqref{eq:potential} with the amplitude $V_0=\lambda M_\text{Pl}^4/(256\pi^2 \xi_h^2)$.

Thus, using potential~\eqref{eq:potential}, we can simultaneously investigate both Starobinsky and Higgs inflationary models. The only free parameter of the potential\,---\,its amplitude $V_0$\,---\,will be used for Bayesian inference. In what follows we will often refer to just Starobinsky potential having in mind both models.

The recent sixth data release from the ACT collaboration unexpectedly challenged the status of the Starobinsky and Higgs inflation models. The obtained value for the spectral index $n_s = 0.9743 \pm 0.0034$ excludes the Starobinsky potential at a $2\sigma$ level of significance \cite{louis2025atacama-980}. In this work, we aim to revive the status of the Starobinsky model by incorporating a non-instantaneous reheating phase.

\subsection{Reheating}

The theory of inflation also requires a stage of post-inflationary reheating, during which the scalar inflaton field decays into Standard Model particles and the stage of the hot Big Bang begins. 

The reheating stage can be modeled using two parameters: the reheating temperature $T_{\text{reh}}$ and the effective parameter of state equation 
\( \bar\omega_{\text{reh}} \), which is defined as follows \cite{ellis2022bicepkeck-6ed}
\begin{equation}
\bar\omega_{\text{reh}} \equiv \frac{1}{N_{\text{reh}} - N_e} \int_{N_e}^{N_{\text{reh}}} \omega_{\text{reh}}(N) dN,
\end{equation}
where \( \omega_{\text{reh}}(N) = \frac{p}{\rho} \) is the parameter in equation of state, linking pressure \(p\)  and energy density \( \rho \), \( N_e \) and \( N_{\text{reh}} \) are the number of $e$-folds at the end of inflation and reheating, respectively.  

Next, to account for the reheating phase in the universe evolution, one needs to calculate the number of $e$-folds from the moment when the pivot mode crosses the horizon to the end of inflation, taking into consideration the reheating stage. This is given by $N_* = \ln\left(\frac{a_\text{end}}{a_*}\right)$, where $a_{\text{end}}$ is the scale factor at the end of inflation and $a_*$ is the scale factor when the pivot mode crosses the horizon. The condition for the mode with momentum $k_{\ast}$ to cross the horizon is defined as $ \frac{k_\ast}{a_\ast H_\ast} =1$, where $H_\ast$ is the Hubble parameter when the pivot mode crosses the horizon. Let us write this expression as follows
\begin{equation}
\frac{k_{\ast}}{a_0}\frac{a_0}{a_{\text{reh}}}  \frac{a_{\text{reh}}}{a_{\text{end}}} \frac{a_{\text{end}}}{a_{\ast}}\frac{1}{H_{\ast}}=1\, .
\end{equation}
where $a$ is the scale factor,
and the lower indices ``0'' and ``reh'' denote quantities calculated today and at the end of reheating, respectively.
According to the entropy conservation law
\begin{equation}
s = g_{*,0}^{(s)} a_0^3T_0^3 = g_{*,\text{reh}} a_{\text{reh}}^3T_{\text{reh}}^3,
\end{equation}
where
$s$ is the entropy density,    
$g_*$ is the effective number of relativistic degrees of freedom,  
$g_*^{(s)}$ is the effective number of entropic degrees of freedom,
$T$ is the temperature.  
Then,
\begin{equation}
\frac{a_0}{a_{\text{reh}}} = \frac{T_{\text{reh}}}{T_0}\Bigg(\frac{g_{*,\text{reh}}}{g_{*,0}^{(s)}}\Bigg)^{\frac{1}{3}}.
\end{equation}
Using the covariant energy conservation law
\begin{equation}
\dot{\rho} + 3H (\rho + p) = 0,
\end{equation}
we can relate the scale factors and energy densities at the end of inflation and reheating through \( \bar\omega_{\text{reh}} \):
\begin{equation}
\frac{a_{\text{reh}}}{a_e} = \left( \frac{\rho_{\text{reh}}}{\rho_e} \right)^{-\frac{1}{3(1 + \bar\omega_{\text{reh}})}}.
\end{equation}

All the energy at the end of the reheating is transferred to ultrarelativistic particles, so the energy density at the end of reheating can be written as
\begin{equation}
\rho_{\text{reh}} = \frac{\pi^2}{30} g_{*,\text{reh}} T_{\text{reh}}^4.
\end{equation}
Then, 
\begin{equation}
\frac{a_{\text{reh}}}{a_e} = \left( \frac{\pi^2}{30} g_{*,\text{reh}}\frac{T_{\text{reh}}^4}{\rho_e} \right)^{-\frac{1}{3(1 + \bar\omega_{\text{reh}})}}.
\end{equation}

Thus, the number of $e$-foldings from the moment when the pivot mode crosses the horizon until the end of inflation is determined by the following expression 
\begin{multline}
N_* = \ln\Bigg[ \frac{H_*}{M_{\text{Pl}}} \frac{a_0M_{Pl}}{k_*} \frac{T_0}{T_{\text{reh}}} \Big(\frac{g_{*,\text{reh}}}{g_{*,0}^{(s)}}\Big)^{\frac{1}{3}} \\ \times\left( \frac{\pi^2}{30} g_{*,\text{reh}}\frac{T_{\text{reh}}^4}{\rho_e} \right)^{\frac{1}{3(1 + \bar\omega_{\text{reh}})}} \Bigg].
\end{multline}

As shown by the authors of Ref.~\cite{martin2024cosmic-687}, for a complete description of the reheating process, it is sufficient to consider a certain combination of the parameters $T_{\text{reh}}$ and $\bar\omega_{\text{reh}}$, rather than each of them separately. Therefore, we will use the so-called rescaled reheating parameter
\begin{multline}
R_{\text{reh}}=\frac{a_e}{a_{\text{reh}}}\left(\frac{\rho_e}{\rho_{\text{reh}}}\right)^\frac{1}{4}\frac{\rho_e^\frac{1}{4}}{M_{\text{Pl}}} \\ = \left( \frac{\pi^2}{30} g_{*,\text{reh}}\frac{T_{\text{reh}}^4}{\rho_e} \right)^{\frac{1}{3(1 + \bar\omega_{\text{reh}})}}\frac{\rho_e^\frac{1}{2}}{\big(\frac{\pi^2}{30} g_{*,\text{reh}}T_{\text{reh}}^4\big)^\frac{1}{4}M_{\text{Pl}}}.
\label{R_reh}
\end{multline}

Finally, using the parameter $R_{\text{reh}}$ and the Friedman equation $\rho_e=\frac{3}{8\pi}H_e^2M_{Pl}^2$, we obtain the expression for determining the required number of $e$-folds
\begin{equation}
N_*=N_0+\ln R_{\text{reh}}+ \ln\left(\frac{H(N_*)}{H_e}\right), 
\label{N_star}
\end{equation}
where
\begin{equation}
N_0 = \ln\left(\frac{T_0a_0}{k_*}\frac{(g_{0,*}^{(s)})^\frac{1}{3}}{g_{\text{reh},*}^\frac{1}{12}}\frac{\sqrt{\pi}}{\sqrt{3}\sqrt[4]{30}}\right) \approx 61.1.
\end{equation}

This expression is an implicit equation for $N_*$, the solution of which depends on the dynamics of the inflaton field. 

\section{Methods}
\label{sec:methods}

We aim to explore the parameter space of the Starobinsky inflationary model \eqref{eq:potential} with non-instantaneous reheating stage using MCMC analysis and the latest ACT DR6 CMB data. In contrast to \cite{martin2024cosmic-687}, where a model-independent approach based on slow-roll parameters is employed, we apply the MCMC algorithm directly to the model parameters $V_0$ and $R_\text{reh}$. Based on the results of our analysis, we compute the Kullback--Leibler divergence to estimate the amount of information that the observational data provide on the reheating phase, specifically the reheating temperature $T_\text{reh}$ and the effective equation of state $\bar\omega_\text{reh}$. Finally, evaluating the Probability to Exceed (PTE), we conclude that the considered model is consistent with current observational data.

\begin{table*}[t]
\renewcommand{\arraystretch}{1.5}
    \centering
\begin{tabular}{@{}c| l |l |l |c@{}}
\hline
\# & \thead{Likelihood} & \thead{Data} & \thead{Multipoles} & \thead{Data \\ points} \\
\hline
\hline
1 & \texttt{planck\_2018\_highl\_plik.TTTEEE\_lite} 
  & \makecell[l]{Foreground-cleaned \\ high-$\ell$ CMB} 
  & \makecell[l]{
    $\ell_\text{TT} = 30\text{--}2600$ \\
    $\ell_\text{TE} = 30\text{--}1996$ \\
    $\ell_\text{EE} = 30\text{--}1996$
    } 
  & 613 \\ \hline
  
2 & \texttt{planck\_2018\_lowl.EE\_sroll2} 
  & Low-$\ell$ CMB 
  & $\ell = 2\text{--}29$ 
  & 28 \\ \hline
  
3 & \texttt{planck\_2018\_lowl.TT} 
  & Low-$\ell$ CMB 
  & $\ell = 2\text{--}29$ 
  & 28 \\ \hline
  
4 & \texttt{planckpr4lensing} 
  & Planck lensing data 
  & $L = 8\text{--}400$ 
  & 9 \\
  \hline
5 & \texttt{act\_dr6\_cmbonly.PlanckActCut} 
  & \makecell[l]{Foreground-cleaned cut \\ Planck high-$\ell$ CMB} 
  & \makecell[l]{
    $\ell_\text{TT} = 30\text{--}1000$ \\
    $\ell_\text{TE} = 30\text{--}600$ \\
    $\ell_\text{EE} = 30\text{--}600$
    } 
  & 225 \\ \hline
  
6 & \texttt{act\_dr6\_cmbonly} 
  & \makecell[l]{Foreground-cleaned \\ high-$\ell$ CMB}  
  & $\ell = 600\text{--}6500$ 
  & 135 \\ \hline
  
7 & \texttt{act\_dr6\_lenslike} 
  & Planck and ACT lensing 
  & $L = 40\text{--}763$ 
  & 19 \\
  \hline
8 & \texttt{bao.desi\_dr2.desi\_bao\_all} 
  & DESI DR2 BAO 
  & -- 
  & 12 \\ \hline
  
9 & \texttt{bicep\_keck\_2018} 
  & CMB B-mode 
  & $\ell = 20\text{--}350$ 
  & 594 \\
\hline
\end{tabular} \\

    \caption{Information about the likelihood functions used in this paper, including a brief description of the data, the range of multipoles studied for CMB-based likelihood functions, and the number of data points. }
    \label{tableLikelihoods}
\end{table*}

It is worth noting that although the parameter $R_\text{reh}$ has no obvious physical meaning, it significantly saves the computational time required for the Bayesian analysis of the reheating model, since it allows us to study one parameter instead of two reheating parameters $T_\text{reh}$ and $\bar\omega_\text{reh}$ which give degenerate predictions for $N_\ast$. Moreover, from the obtained posterior distribution of the parameter $R_\text{reh}$, we can obtain the posterior distributions of the physical parameters $T_\text{reh}$, $\bar\omega_\text{reh}$ using the computationally much less expensive MCMC algorithm, the details of which we will discuss in Appendix~\ref{app:MCMC}.

To implement the Bayesian inference, we use the Boltzmann code \texttt{CLASS} \cite{blas2011cosmic-8f1, lesgourgues2011cosmic-c31}, which solves all the background dynamics of the Universe evolution, and the \texttt{cobaya} software \cite{torrado2020cobaya-7d4}, which implements the MCMC algorithm. The listing of \texttt{CLASS} settings used in the \texttt{cobaya} runs is presented in Appendix~\ref{app:classy_config}. We analyze the resulting chains and plot them using the capabilities of the \texttt{GetDist} package \cite{lewis2019getdist-ed7}. 

By default, \texttt{CLASS} does not account for the reheating stage. To address this, we modified the \texttt{CLASS} code by implementing the reheating stage in the \texttt{primordial} module. Specifically, by numerically solving Eq.~\eqref{N_star} using the bisection method, \texttt{CLASS} now calculates the number of $e$-foldings from the end of inflation when the pivot mode crosses the horizon for the given parameters $R_\text{reh}$ and $V_0$.

In the base Bayesian analysis, we use data on the angular spectra of temperature (TT), polarization (TE, EE, TB, EB, BB) and lensing ($\phi \phi$) contained in the likelihood functions \texttt{planck\_2018\_lowl.EE\_sroll2}, \texttt{planck\_2018\_lowl.TT}~\cite{collaboration2019planck-584}, \texttt{act\_dr6\_cmbonly}, \texttt{act\_dr6\_} \texttt{mbonly.PlanckActCut},  \mbox{\texttt{act\_dr6\_lenslike}~\cite{louis2025atacama-980}} and \mbox{\texttt{bicep\_keck\_2018}~\cite{collaboration2021improved-f44}}, as well as the BAO observations contained in the likelihood function \texttt{bao.desi\_dr2.desi\_bao\_all}~\cite{collaboration2025desi-e55}~\footnote{We also performed an analysis using the DESI Year One data, included in the \texttt{bao.desi\_2024\_bao\_all}~\cite{collaboration2024desi-2ba, collaboration2024desi-020, collaboration2024desi-265} likelihood function, as done by the ACT collaboration. We obtained the same results, independent of the data release used.} 

We also perform the Bayesian inference without the ACT collaboration data for comparative analysis. Accordingly, we used in this analysis the foreground-marginalized likelihood function \texttt{planck\_2018\_highl\_plik.TTTEEE\_lite}, which contains the CMB high-$\ell$ data, and \texttt{planckpr4lensing}~\cite{carron2022cmb-ee4}, which contains the lensing data. Detailed information about these functions can be found in Table \ref{tableLikelihoods}, including the number of data points, measured multipoles etc. 

In the presentation of the results, we use the notation agreed upon by the ACT collaboration, i.e., the abbreviation ``Planck-LB-BK18'' will represent the combination of measurements 1--4, 8, 9, and ``P-ACT-LB-BK18'' will represent the combination of measurements 2, 3, 5--9, respectively (see Table~\ref{tableLikelihoods} for the numbering).

As the prior distributions, we take homogeneous distributions on the logarithmic scale, including a wide range of values of $\ln R_{\text{reh}}\in[-46, 15+\frac{1}{3}\ln \rho_\text{end}]$ and $\ln (V_0/M_\text{Pl}^4)\in[-150,-28]$ in Planck masses. These values are due to the fact that the reheating that occurs after inflation requires that $\rho_{\text{reh}} \leq \rho_{\text{end}}$, while we expect the average equation of state of the Universe during this stage to satisfy the condition $-1/3 < \bar\omega_{\text{reh}} < 1$, where the lower bound guarantees that inflation has stopped and the upper bound follows from causality restrictions. The requirement to avoid disrupting Big Bang nucleosynthesis imposes the constraint $\rho_{\text{reh}} > \rho_{\text{nuc}}$, and we set the lower bound at $\rho_{\text{nuc}}=g_{\ast,\text{nuc}}\frac{\pi^2}{30} T_{\text{nuc}}^4$ with $T_{\text{nuc}}=10\,$MeV. Equation~\eqref{R_reh} and the condition $\rho_{\text{end}} < M_\text{Pl}^4$ give the widest possible prior distribution for the reheating parameter $\ln R_{\text{reh}}=[-46, 15+\frac{1}{3}\ln \rho_\text{end}]$~\cite{martin2024cosmic-687}. The choice of the upper limit for $\ln (V_0/M_\text{Pl}^4)=-28$ is due to the fact that at higher values of the parameter, the agreement with the Planck Collaboration data becomes impossible.

Such limits $\ln R_\text{reh}$ correspond to the limits $\ln T_\text{reh}/M_\text{Pl} \in [-48.5, -8.5]$ and $\bar\omega_\text{reh} \in [-1/3, 1]$ on the temperature and the effective equation of state of the reheating, respectively. We will need these values to analyze the parameter space $(\ln \tfrac{T_\text{reh}}{M_\text{Pl}},\,\bar\omega_\text{reh},\,\ln \frac{V_0}{M_\text{Pl}^4} )$ based on the two-dimensional posterior distribution of the parameters $(\ln R_\text{reh},\,\ln \frac{V_0}{M_\text{Pl}^4})$ that will be obtained by base Bayesian inference.

For all the necessary `nuisance' parameters, we choose standard prior distributions given in the files of the corresponding likelihood functions in \texttt{cobaya}.

As a result of the Bayesian inference, we obtain the posterior distributions of model parameters, which allow us to calculate the information gain that can be obtained from the observational data compared to the prior information. To do this, we will use the Kullback--Leibler divergence~\cite{kullback1951information-d27}:
\begin{equation}
D_{\text{KL}} = \int P(\theta | D) \log \frac{P(\theta| D)}{P(\theta)} \, d\theta.
\label{DKL}
\end{equation}
This value is a measure of how much the obtained posterior distribution differs from the corresponding prior distribution.

Finally, we evaluate the model fit to the data using Probability to Exceed (PTE). PTE is the probability that a random variable $\chi^2$ exceeds the observed value if the model is correct, i.e.
\begin{equation}
    \text{PTE} = P(\chi^2 > \chi^2_{\text{obs}} \mid \text{dof}),
\end{equation}
where $\chi^2_{\text{obs}}$ is the observed value of the statistic $\chi^2$, and dof is the number of degrees of freedom. $\text{PTE} \approx 0.5$ indicates a good fit between the model and the data. A PTE value close to zero indicates that the model fails to explain the data (i.e., the observed $\chi^2$ is too large), while a value close to one may suggest overfitting or an overestimation of the observational uncertainties. In Python, the PTE value can be computed using the \texttt{chi2.sf} function from the \texttt{scipy.stats}~\cite{SciPy2020} module as \texttt{chi2.sf(chi2\_obs, dof)}. The PTE is commonly used as an indicator of the goodness of fit between the model and the data in Bayesian analysis.

\section{Results and discussion}
\label{sec:results}

\begin{figure}[t]
    \centering
    \includegraphics[width=\columnwidth]{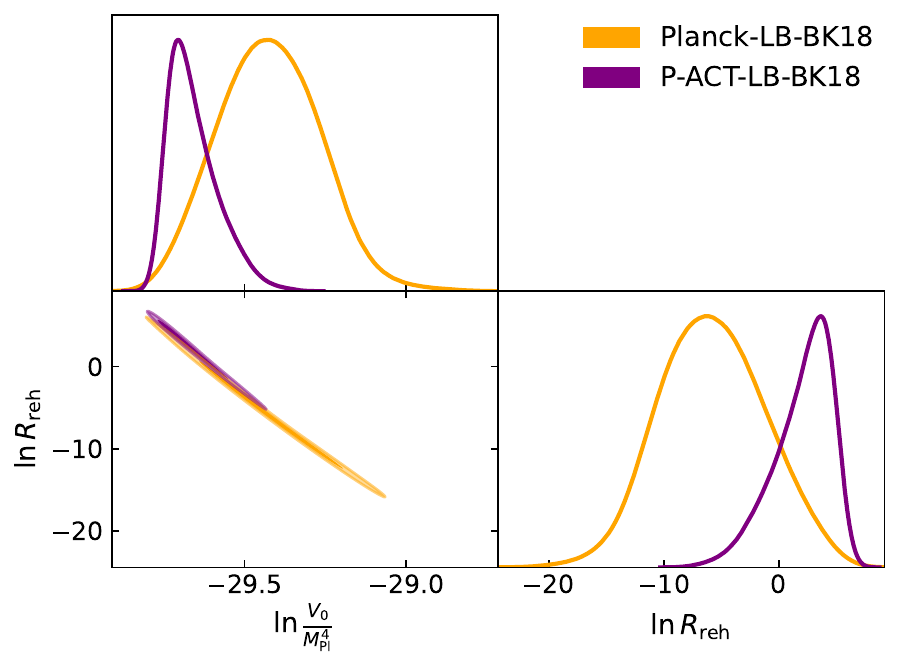}
    \caption{The two-dimensional posterior distribution for the parameters $\ln R_{\text{reh}}$ and $\ln (V_0/M_\text{Pl}^4)$ obtained by statistical analysis of the Planck-LB-BK18 and P-ACT-LB-BK18 data. The credible limits of $95\%$ and $68\%$ are indicated.}
    \label{fig:R_V_cornerplot}
\end{figure}

As a result of our analysis of Markov chains with an effective size of $\gtrsim 10^4$ points, we obtained posterior distributions for the parameters $\ln R_{\text{reh}}$ and $\ln (V_0/M_\text{Pl}^4)$, which are shown in Fig.~\ref{fig:R_V_cornerplot}. Also, the best-fit values, mean values with standard deviation, and $95\%$ credible limits for $\ln R_{\text{reh}}$, $\ln (V_0/M_\text{Pl}^4)$ and supplementary derived parameters are shown in Table \ref{table1}.

\begin{table}[t]\renewcommand{\arraystretch}{1.3}
    \centering
\begin{tabular}{c|c|c|c|cc} 
\hline
Param & Best-fit & Mean$\pm\sigma$ & 95\% lower & 95\% upper \\ \hline 
\hline
$\ln R_\text{reh}$ & $4.302$ & $2.124_{-2.6}^{+2.5}$ & $-3.98$ & $5.32$ \\ \hline
$\ln \frac{V_0}{M_\text{Pl}}$ & $29.729$ & $29.661_{-0.08}^{+0.08}$ & $29.8$ & $29.5$ \\ \hline
$n_\text{s}$ & $0.971$ & $0.970_{-0.0012}^{+0.0011}$ & $0.967$ & $0.972$ \\ \hline
$\ln 10^{10}A_\text{s}$ & $3.063$ & $3.065_{-0.01}^{+0.01}$ & $3.05$ & $3.08$ \\ \hline
$r$ & $0.0024$ & $0.0026_{-0.9679}^{+0.00}$ & $0.00233$ & $0.00309$ \\ \hline
$N_\ast$ & $66.1$ & $64.0_{-2.6}^{+2.5}$ & $57.9$ & $67.2$ \\ \hline
\end{tabular} \\

    \caption{Best-fit values, means with standard deviation and 95\% limits for $\ln R_{\text{reh}}$, $\ln (V_0/M_\text{Pl}^4)$ and some of the derived parameters of interest obtained by the combined analysis of P-ACT-LB-BK18. The joint likelihood function was maximized to $-\ln{\cal L}_\mathrm{min} =682.52$, which corresponds to the minimum value for the function $\chi^2=1365.04$.}
    \label{table1}
\end{table}

\begin{table}[t]
\resizebox{0.45\textwidth}{!}{
\renewcommand{\arraystretch}{1.3}
\centering
\begin{tabular}{c|c|c|c|c|c|c|c} 
\hline
$\chi^2_\text{P-ACT-LB-BK18}$ & $\chi^2_1$ & $\chi^2_2$ & $\chi^2_5$ & $\chi^2_6$ & $\chi^2_7$ & $\chi^2_8$ & $\chi^2_9$ \\ \hline \hline
$1365.04$ & $391.28$ & $22.47$ & $220.50$ & $159.29$ & $19.55$ & $14.71$ & $537.25$ \\ \hline
\end{tabular} }\\
    \caption{Best-fit values of $\chi^2$ corresponding to the likelihood functions used for P-ACT-LB-BK18 analysis. The index corresponds to the numbering from the table \ref{tableLikelihoods}.}
    \label{table_chi}
\end{table}

\begin{figure}[t]
    \centering
    \includegraphics[width=\columnwidth]{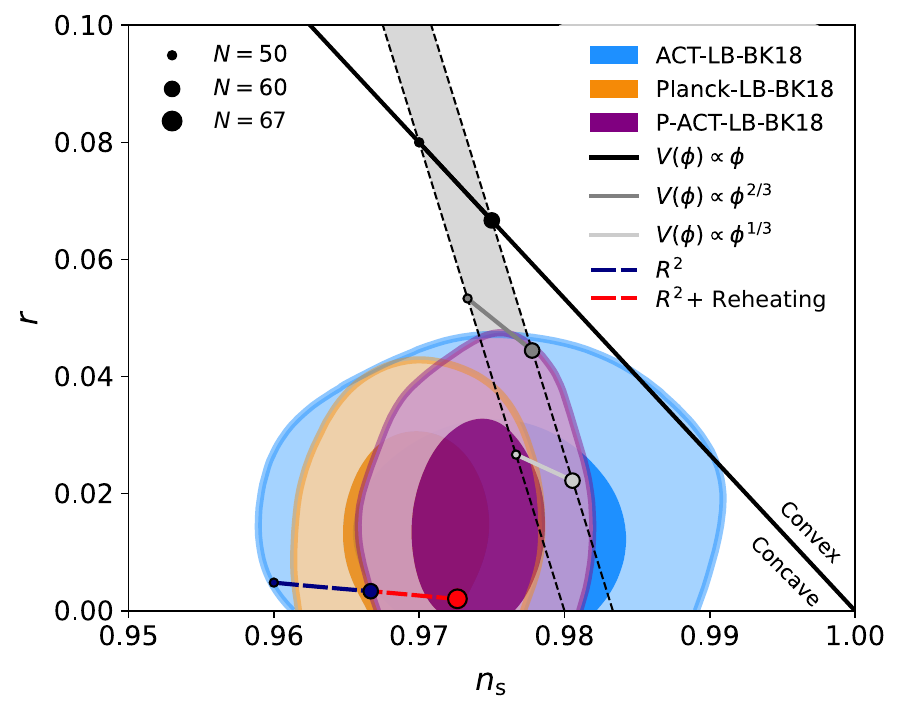}
    \caption{Constraints on the scalar and tensor primordial power spectra at the pivot scale $k_* = 0.05~\mathrm{Mpc}^{-1}$, shown in the $r$--$n_s$ parameter space. The constraints on $r$ are driven by the BK18 data, while the constraints on $n_s$ are driven by \textcolor{orange}{Planck} (orange), \textcolor{blue}{ACT} (blue), or \textcolor{purple}{P-ACT} (purple). The combined dataset also includes CMB lensing and BAO in all cases. The various circles and solid lines within the gray band show predictions for different power-law potentials with the number of $e$-folds of inflation $50 < N < 60$. The Starobinsky $R^2$ model is also shown for $50$--$60$ $e$-folds (dashed navy line) and for $60$--$67$ $e$-folds (dashed red line) reflecting the impact of the reheating phase on the number of $e$-folds. Figure adapted from the ACT collaboration~\cite{calabrese2025atacama-a33}.}
    \label{fig:n_s_r_cornerplot}
\end{figure}

\begin{table}[t]\renewcommand{\arraystretch}{1.3}
    \centering

\begin{tabular}{l|c|c|c|c} 
\hline
Param & Best-fit & Mean$\pm\sigma$ & 95\% lower & 95\% upper \\ \hline \hline
$\ln \frac{T_\text{reh}}{M_\text{Pl}}$ & $-43.9558$ & $-41.597_{-5.9}^{+6.3}$ & $-48.4$ & $-24.8$ \\ \hline
$\bar\omega_\text{reh}$ & $0.9608$ & $0.881_{-0.1}^{+0.1}$ & $0.616$ & $0.997$ \\ \hline
$\ln \frac{V_0}{M_\text{Pl}^4}$ & $-29.70$ & $-29.65_{-0.07}^{+0.08}$ & $-29.8$ & $-29.5$ \\ \hline
\end{tabular} \\
    \caption{Best-fit values, means, standard deviations, and $95\%$ limits for the parameters $\ln T_{\text{reh}}/M_\text{Pl}$, $\bar\omega_{\text{reh}}$, and $\ln (V_0/M_{\text{Pl}}^4)$ obtained by Bayesian inference using the P-ACT-LB-BK18 data.}
    \label{table2}
\end{table}

The table \ref{table_chi} shows the best-fit values of the $\chi^2$ functions for the combined dataset and for each dataset separately. The PTE calculation does not take into account the $\chi^2$ for the likelihood function \texttt{planck\_2018\_lowl.EE\_sroll2}, since this function is not a classical $\chi^2$, but instead it is compared to the reference value provided by the Planck calibration, $\chi_\text{lowl.EE}^2 \approx 396$. Thus, for the best-fit analysis, the value of $\chi^2_\text{P-ACT-LB-BK18} = 973. 76$ without taking into account $\chi_1^2$ for 1029 degrees of freedom (1040 data points and 11 parameters), we obtain $\text{PTE}=89\%$, which indicates a very high degree of consistency of the Starobinsky inflation model with non-instantaneous reheating with the most recent observational data P-ACT-LB-BK18.

\begin{figure*}[t]
    \centering
    \includegraphics[width=0.75\textwidth]{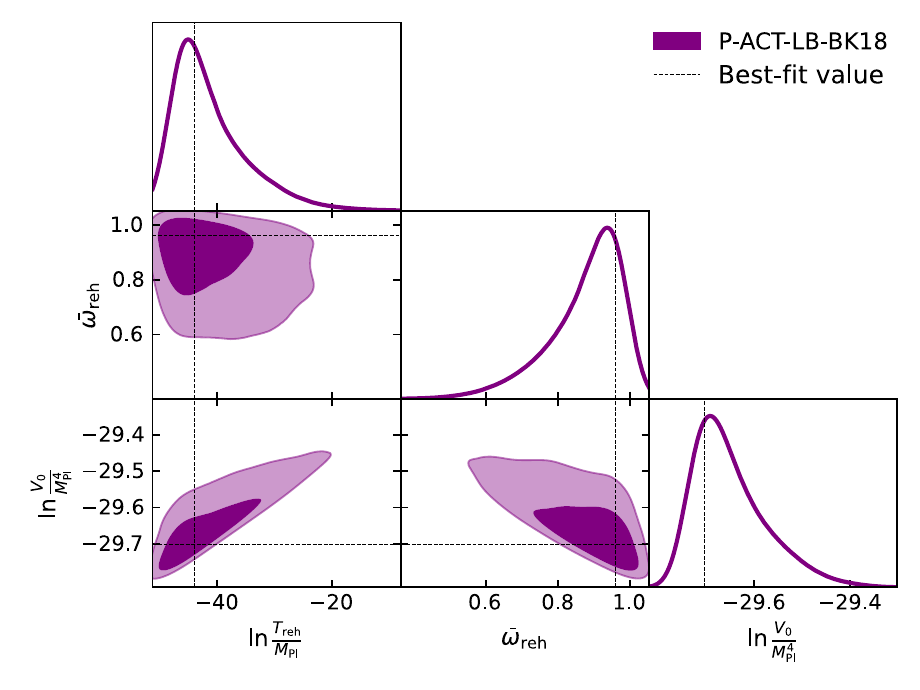}
    \caption{Two-dimensional posterior distribution for the parameters $\ln T_{\text{reh}}/M_\text{Pl}$, $\bar\omega_{\text{reh}}$, and $\ln (V_0/M_{\text{Pl}}^4)$ obtained by Bayesian inference of P-ACT-LB-BK18 data. The credible limits of $95\%$ and $68\%$ are circled. The best-fit values of the parameters are represented by dashed lines.}
    \label{fig:T_omega_V_cornerplot}
\end{figure*}

The ACT collaboration provides a spectral index value of $n_\text{s}=0.9743 \pm 0.0034$~\cite{louis2025atacama-980} for the pivot mode on the CMB scale with momentum $k_\ast = 0. 05 \, \text{Mpc}^{-1}$, which excludes Starobinsky inflation with $2\sigma$ level if the pivot mode goes beyond the horizon for $N_\ast = 50\text{--}60$ $e$-foldings before the end of the inflation. However, from our analysis, we find that non-instantaneous reheating can extend the postinflationary expansion and, therefore, postpone the horizon exit of the CMB pivot scale to $N_\ast = 67$ $e$-folds from the end of inflation which gives the value of the scalar spectral index $n_\text{s}=0.972$ (see Figure~\ref{fig:n_s_r_cornerplot}). Thus, taking into account non-instantaneous reheating, Starobinsky inflation can be still considered as a viable inflationary model.

Also, from Table~\ref{table1} we can conclude that the observational data in combination of Planck, ACT and BAO provide quite strict constraints on the inflation parameter $\ln (V_0/M_\text{Pl}^4)$, while the constraints on the reheating parameter $\ln R_{\text{reh}}$ are less strict: $R_{\text{reh}}$ can take values within $4$ orders of magnitude. Although still the $95\%$ credible interval is wide, it is already much more restricted than when analyzing only Planck and BAO data without ACT. 

The information gain that observational data provide about the model parameters can be quantified using the Kullback--Leibler divergence \eqref{DKL}:
\begin{gather}
D_{\text{KL}}^{R_\text{reh}} = \int P(\ln R_{\text{reh}} | D) \log \frac{P(\ln R_{\text{reh}} | D)}{P(\ln R_{\text{reh}})} \, d\ln R_{\text{reh}}, \\
D_{\text{KL}}^{\text{inf}} = \int P(\ln \tfrac{V_0}{M_\text{Pl}^4} | D) \log \frac{P(\ln \tfrac{V_0}{M_\text{Pl}^4} | D)}{P(\ln \tfrac{V_0}{M_\text{Pl}^4} )} \, d\ln \tfrac{V_0}{M_\text{Pl}^4}.
\end{gather}

For our model, we calculated the values of $D_{\text{KL}}^{R_\text{reh}} \approx 2.52$ bits, $D_{\text{KL}}^{\text{inf}} \approx 8.66$ bits for P-ACT-LB-BK18 analysis and $D_{\text{KL}}^{R_\text{reh}} \approx 1.44$ bits, $D_{\text{KL}}^{\text{inf}} \approx 7.53$ bits for the Planck-LB-BK18 analysis. Thus, the data published by the ACT collaboration yields an additional $1.08$ bits of information about the reheating stage, which is an increase of $75\%$. The amount of information that we can obtain from the observational data on the reheating stage of $2.52$ bits is already large enough to conclude that it is necessary to take into account the model of non-instantaneous reheating when considering the evolution of the Universe. Interpretation of the data without taking into account the reheating can distort the real picture of the evolution of the Universe. 

From the obtained two-dimensional posterior distribution of the parameters $\ln R_{\text{reh}}$, $\ln (V_0/M_{\text{Pl}}^4)$, we can calculate the distributions for the parameters $\ln T_{\text{reh}}/M_\text{Pl}$ and $\bar\omega_{\text{reh}}$ by performing a much less computationally expensive MCMC algorithm, the details of which are described in Appendix A. The results of this analysis are presented in Table~\ref{table2} and Figure~\ref{fig:T_omega_V_cornerplot}.

By calculating the Kullback--Leibler divergence \eqref{DKL} for the two-dimensional posterior distribution of the parameters $\theta = (\ln \tfrac{T_\text{reh}}{M_\text{Pl}}, \, \bar\omega_{\text{reh}})$, we obtained $D_{\text{KL}}^{\text{reh}} \approx 2.92$, as well as for the one-dimensional posterior distributions for temperature $D_{\text{KL}}^{T_\text{reh}} \approx 0.95$, and for the effective equation of state $D_{\text{KL}}^{\bar\omega_\text{reh}} \approx 1.95$. This indicates that in the parameterization $\theta = (\ln \tfrac{T_\text{reh}}{M_\text{Pl}}, \, \bar\omega_{\text{reh}})$ the amount of information about the reheating stage that we can obtain from the observational data is even greater.

\section{Conclusions}
\label{sec:conclusion}

We performed a Bayesian analysis of the Starobinsky and Higgs inflationary models with a non-instantaneous reheating phase using the combined dataset, which includes the latest observations on CMB from Planck 2018 release, ACT DR6 and BICEP/Keck 2018 and on BAO from DESI DR2. Markov Chain Monte Carlo analysis, based on chains with an effective sample size of over $10^4$, yielded robust constraints on the inflationary energy scale $V_0$, the reheating parameter $R_\text{reh}$, and derived quantities related to the reheating phase.

The results show that the inflationary energy scale $V_0$ is tightly constrained with sub-percent level precision, while the constraints on $R_{\text{reh}}$ remain broader, spanning approximately four orders of magnitude. However, the inclusion of ACT data improves the constraint on reheating by increasing the information gain, measured via the Kullback--Leibler divergence, by 75\%, reaching $D_{\text{KL}}^{R_{\text{reh}}} \approx 2.52$ bits. This demonstrates that the reheating phase, though still weakly constrained, is significantly informed by current data and should not be neglected in cosmological analyses.

The extended analysis of the reheating parameters $T_{\text{reh}}$ and $\bar\omega_{\text{reh}}$, obtained via reparametrization from the two-dimensional posterior distribution of $(R_{\text{reh}}, V_0)$, reveals that the effective equation of state during reheating is constrained to values close to a stiff regime, with $\bar\omega_{\text{reh}} \approx 0.88 \pm 0.1$, and the reheating temperature average is approximately $T_\text{reh}\approx11.56~\text{GeV}$, which means a sufficiently long reheating stage. The total information gain on reheating in this parametrization reaches $D_{\text{KL}}^{\text{reh}} \approx 2.92$ bits, with the majority attributed to constraints on $\bar\omega_{\text{reh}}$.

The present study highlights the importance of including realistic reheating dynamics when connecting inflationary models to CMB data and reinforces the necessity of precise modeling of the post-inflationary universe in future cosmological studies. Results of this work show that the Starobinsky or Higgs inflationary models with non-instantaneous reheating may still be viable in light of current observations. In particular, allowing for a delayed reheating phase naturally extends the number of $e$-folds to $N_\ast \approx 67$, resolving the mild tension between the predicted and observed scalar spectral index.
Previous studies of the universe reheating after Starobinsky~\cite{Gorbunov:2010bn,Jeong:2023zrv} and Higgs~\cite{Bezrukov:2008ut,Garcia-Bellido:2008ycs,Bezrukov:2011sz,Ema:2016dny,DeCross:2016cbs} inflation, although can reach moderately low reheating temperatures of order $10^8\,$GeV, typically suggest equation of state close to zero. Therefore, in order to achieve the stiff equation of state during reheating one may need to consider extra ingredients making the inflaton potential very flat close to the origin, like in models with post-inflationary kination stage~\cite{Turner:1983he,Joyce:1996cp}. We leave this interesting possibility for future studies.

\bigskip
\textbf{Note added.} While finalizing our research we found a recent work~\cite{Drees:2025ngb} on the similar topic. The authors did analytical computations of inflationary predictions for $n_s$ and $r$ without performing the MCMC analysis and arrive at similar conclusions to ours, giving a preference to low reheating temperatures and stiff equation of state. Moreover, soon after our work, another two articles~\cite{Liu:2025qca,Haque:2025uis} appeared reporting similar results. Finally, we would like to mention a number of recent articles~\cite{Kallosh:2025rni,Aoki:2025wld,Gialamas:2025kef,Dioguardi:2025vci,Salvio:2025izr,Antoniadis:2025pfa,He:2025bli,Kuralkar:2025zxr,Yogesh:2025wak,Gialamas:2025ofz,Yin:2025rrs} inspired by the ACT DR6 proposing different inflationary scenarios allowing for a slightly increased value of $n_s$.

\subsection*{Acknowledgements}

The work of O.~O.~S. is sustained by a Philipp--Schwartz fellowship of the University of M\"{u}nster. The work of S.~I.~V. is supported by a Swiss National Science Foundation, grant extension IZSEZO.216773 and  University of Geneva grant for Researchers at Risk.

Based on observations obtained with Planck (\href{http://www.esa.int/Planck}{http://www.esa.int/Planck}), an ESA science mission with instruments and contributions directly funded by ESA Member States, NASA, and Canada.

This study used data obtained with the BICEP/Keck Collaboration and the Atacama Cosmology Telescope Collaboration.

This research used data obtained with the Dark Energy Spectroscopic Instrument (DESI). DESI construction and operations is managed by the Lawrence Berkeley National Laboratory. This material is based upon work supported by the U.S. Department of Energy, Office of Science, Office of High-Energy Physics, under Contract No. DE–AC02–05CH11231, and by the National Energy Research Scientific Computing Center, a DOE Office of Science User Facility under the same contract. Additional support for DESI was provided by the U.S. National Science Foundation (NSF), Division of Astronomical Sciences under Contract No. AST-0950945 to the NSF’s National Optical-Infrared Astronomy Research Laboratory; the Science and Technology Facilities Council of the United Kingdom; the Gordon and Betty Moore Foundation; the Heising-Simons Foundation; the French Alternative Energies and Atomic Energy Commission (CEA); the National Council of Humanities, Science and Technology of Mexico (CONAHCYT); the Ministry of Science and Innovation of Spain (MICINN), and by the DESI Member Institutions: www.desi.lbl.gov/collaborating-institutions. The DESI collaboration is honored to be permitted to conduct scientific research on I’oligam Du’ag (Kitt Peak), a mountain with particular significance to the Tohono O’odham Nation. Any opinions, findings, and conclusions or recommendations expressed in this material are those of the author(s) and do not necessarily reflect the views of the U.S. National Science Foundation, the U.S. Department of Energy, or any of the listed funding agencies.

We use the Boltzmann code \texttt{CLASS} \cite{blas2011cosmic-8f1, lesgourgues2011cosmic-c31}, the \texttt{cobaya} software \cite{torrado2020cobaya-7d4}, the \texttt{GetDist}~\cite{lewis2019getdist-ed7} and the \texttt{scipy.stats}~\cite{SciPy2020} packages. 

\appendix

\section{MCMC analysis for reheating parameters}
\label{app:MCMC}

This appendix provides a brief overview of the MCMC algorithm used to obtain the posterior distributions of the parameters $\ln T_\text{reh}/M_\text{Pl}$ and $\bar\omega_\text{reh}$ from the two-dimensional posterior distribution of the parameters $\ln R_\text{reh}$ and $\ln (V_0/M_\text{Pl}^4)$. In essence, we aim to perform a reparameterization of the parameter space from $\theta = (\ln R_\text{reh},\ln \tfrac{V_0}{M_\text{Pl}^4})$ to $\theta' = (\ln \tfrac{T_\text{reh}}{M_\text{Pl}}, \bar\omega_\text{reh},\ln \tfrac{V_0}{M_\text{Pl}^4})$. However, since we do not possess an analytical expression for $T_\text{reh}$ and $\bar\omega_\text{reh}$ as functions of $R_\text{reh}$, a full MCMC analysis must be conducted in the extended parameter space $\theta'$, using as data the two-dimensional posterior distribution of the original parameters $\theta$. In this case, the likelihood function is defined as
\begin{multline}
\ln\mathcal{L}(\ln \tfrac{T_\text{reh}}{M_\text{Pl}}, \bar\omega_\text{reh}, \ln \tfrac{V_0}{M_\text{Pl}^4}) \\ = \ln P\big(\ln R_\text{reh}\left(\ln \tfrac{T_\text{reh}}{M_\text{Pl}}, \bar\omega_\text{reh}, \ln \tfrac{V_0}{M_\text{Pl}^4}\right), \ln \tfrac{V_0}{M_\text{Pl}^4}\big)\,,
\end{multline} 
where $P(\ln R_\text{reh},\ln \tfrac{V_0}{M_\text{Pl}^4})$ is the two-dimensional posterior distribution obtained from the Bayesian analysis of the P-ACT-LB-BK18 data.

Moreover, for each step in the Markov chains, one needs to compute the value of the parameter $R_\text{reh}$ in the parameter space $\theta$. Since, by definition in Eq.~\eqref{R_reh}, $R_\text{reh}$ depends on the energy density at the end of inflation $\rho_\text{end}$, it is necessary to solve the background equations at each step. This task is handled by the modified Boltzmann code \texttt{CLASS}. However, there is no need to solve the mode equations or compute the primordial scalar power spectrum, which significantly reduces the computational time.

The analysis employs broad logarithmic flat priors $\ln T_\text{reh}/M_\text{Pl} \in [-48.5, -8.5]$ and $\bar\omega_\text{reh} \in [-1/3, 1]$ for the reheating temperature and the effective equation of state, respectively.

\section{\texttt{CLASS} configuration for \texttt{cobaya} run}
\label{app:classy_config}

\begin{figure}[!htb]
    \centering
    \begin{lstlisting}
    extra_args:
      k_pivot: 0.05
      h: 0.6822
      omega_b: 0.02256
      omega_cdm: 0.1179
      tau_reio: 0.0632
      N_ncdm: 1
      m_ncdm: 0.06
      N_ur: 2.0308
      T_cmb: 2.7255
      YHe: BBN
      non_linear: hmcode
      hmcode_version: '2020'
      recombination: HyRec
      lensing: 'yes'
      output: lCl , tCl , pCl , mPk
      modes: s, t
      l_max_scalars: 9500
      delta_l_max: 1800
      P_k_max_h/Mpc: 100.
      l_logstep: 1.025
      l_linstep: 20
      perturbations_sampling_stepsize : 0.05
      l_switch_limber: 30.
      hyper_sampling_flat: 32.
      l_max_g: 40
      l_max_ur: 35
      l_max_pol_g: 60
      ur_fluid_approximation: 2
      ur_fluid_trigger_tau_over_tau_k: 130.
      radiation_streaming_approximation: 2
      radiation_streaming_trigger_tau_over_tau_k: 240.
      hyper_flat_approximation_nu: 7000.
      transfer_neglect_delta_k_S_t0: 0.17
      transfer_neglect_delta_k_S_t1: 0.05
      transfer_neglect_delta_k_S_t2: 0.17
      transfer_neglect_delta_k_S_e: 0.17
      accurate_lensing: 1
      start_small_k_at_tau_c_over_tau_h: 0.0004
      start_large_k_at_tau_h_over_tau_k: 0.05
      tight_coupling_trigger_tau_c_over_tau_h: 0.005
      tight_coupling_trigger_tau_c_over_tau_k: 0.008
      start_sources_at_tau_c_over_tau_h: 0.006
      l_max_ncdm: 30
      tol_ncdm_synchronous: 1e-6
      hmcode_max_k_extra: 0
    \end{lstlisting}
    \caption{Baseline settings used for \texttt{CLASS} theory calculations, updating the default assumptions of the public \texttt{CLASS} version. In this work, we use a version of the code that has been updated to implement the latest \texttt{HMcode-2020} model for the non-linear power spectrum, provided by J.~Lesgourgues (developed from \texttt{CLASS} v3.2.2).}
    \label{listing}
\end{figure}

The configuration file used for the \texttt{CLASS} runs is given in Fig.~\ref{listing}. All parameters are the same as those used by the ACT collaboration, with the baseline $\Lambda$CDM parameters set to the mean values from the ACT results. An important line `\texttt{hmcode\_max\_k\_extra: 0}' was added, since in the analysis of inflationary models it is necessary to numerically compute the primordial scalar perturbations spectrum, and the parameter \texttt{hmcode\_max\_k\_extra} $> 0$, which considers the spectrum on extra modes, is only compatible with methods that use an analytical spectrum.

\bibliographystyle{apsrev4-2.bst}
\bibliography{references.bib}

\begin{thebibliography}{68}%
\makeatletter
\providecommand \@ifxundefined [1]{%
 \@ifx{#1\undefined}
}%
\providecommand \@ifnum [1]{%
 \ifnum #1\expandafter \@firstoftwo
 \else \expandafter \@secondoftwo
 \fi
}%
\providecommand \@ifx [1]{%
 \ifx #1\expandafter \@firstoftwo
 \else \expandafter \@secondoftwo
 \fi
}%
\providecommand \natexlab [1]{#1}%
\providecommand \enquote  [1]{``#1''}%
\providecommand \bibnamefont  [1]{#1}%
\providecommand \bibfnamefont [1]{#1}%
\providecommand \citenamefont [1]{#1}%
\providecommand \href@noop [0]{\@secondoftwo}%
\providecommand \href [0]{\begingroup \@sanitize@url \@href}%
\providecommand \@href[1]{\@@startlink{#1}\@@href}%
\providecommand \@@href[1]{\endgroup#1\@@endlink}%
\providecommand \@sanitize@url [0]{\catcode `\\12\catcode `\$12\catcode `\&12\catcode `\#12\catcode `\^12\catcode `\_12\catcode `\%12\relax}%
\providecommand \@@startlink[1]{}%
\providecommand \@@endlink[0]{}%
\providecommand \url  [0]{\begingroup\@sanitize@url \@url }%
\providecommand \@url [1]{\endgroup\@href {#1}{\urlprefix }}%
\providecommand \urlprefix  [0]{URL }%
\providecommand \Eprint [0]{\href }%
\providecommand \doibase [0]{https://doi.org/}%
\providecommand \selectlanguage [0]{\@gobble}%
\providecommand \bibinfo  [0]{\@secondoftwo}%
\providecommand \bibfield  [0]{\@secondoftwo}%
\providecommand \translation [1]{[#1]}%
\providecommand \BibitemOpen [0]{}%
\providecommand \bibitemStop [0]{}%
\providecommand \bibitemNoStop [0]{.\EOS\space}%
\providecommand \EOS [0]{\spacefactor3000\relax}%
\providecommand \BibitemShut  [1]{\csname bibitem#1\endcsname}%
\let\auto@bib@innerbib\@empty
%</preamble>
\bibitem [{\citenamefont {Weinberg}(2008)}]{Weinberg:2008zzc}%
  \BibitemOpen
  \bibfield  {author} {\bibinfo {author} {\bibfnamefont {S.}~\bibnamefont {Weinberg}},\ }\href@noop {} {\emph {\bibinfo {title} {Cosmology}}}\ (\bibinfo  {publisher} {{Oxford University Press}},\ \bibinfo {address} {New York},\ \bibinfo {year} {2008})\BibitemShut {NoStop}%
\bibitem [{\citenamefont {Dodelson}\ and\ \citenamefont {Schmidt}(2021)}]{dodelsonModernCosmology2021}%
  \BibitemOpen
  \bibfield  {author} {\bibinfo {author} {\bibfnamefont {S.}~\bibnamefont {Dodelson}}\ and\ \bibinfo {author} {\bibfnamefont {F.}~\bibnamefont {Schmidt}},\ }\href@noop {} {\emph {\bibinfo {title} {Modern cosmology}}},\ \bibinfo {edition} {second edition}\ ed.\ (\bibinfo  {publisher} {Academic Press},\ \bibinfo {address} {London, United Kingdom ; San Diego, CA},\ \bibinfo {year} {2021})\BibitemShut {NoStop}%
\bibitem [{\citenamefont {Smoot}\ \emph {et~al.}(1992)\citenamefont {Smoot} \emph {et~al.}}]{smoot1992structure-837}%
  \BibitemOpen
  \bibfield  {author} {\bibinfo {author} {\bibfnamefont {G.~F.}\ \bibnamefont {Smoot}} \emph {et~al.} (\bibinfo {collaboration} {COBE}),\ }\bibfield  {title} {\bibinfo {title} {{Structure in the COBE differential microwave radiometer first year maps}},\ }\href {https://doi.org/10.1086/186504} {\bibfield  {journal} {\bibinfo  {journal} {Astrophys. J. Lett.}\ }\textbf {\bibinfo {volume} {396}},\ \bibinfo {pages} {L1} (\bibinfo {year} {1992})}\BibitemShut {NoStop}%
\bibitem [{\citenamefont {Spergel}\ \emph {et~al.}(2003)\citenamefont {Spergel} \emph {et~al.}}]{spergel2003firstyear-505}%
  \BibitemOpen
  \bibfield  {author} {\bibinfo {author} {\bibfnamefont {D.~N.}\ \bibnamefont {Spergel}} \emph {et~al.} (\bibinfo {collaboration} {WMAP}),\ }\bibfield  {title} {\bibinfo {title} {{First year Wilkinson Microwave Anisotropy Probe (WMAP) observations: Determination of cosmological parameters}},\ }\href {https://doi.org/10.1086/377226} {\bibfield  {journal} {\bibinfo  {journal} {Astrophys. J. Suppl.}\ }\textbf {\bibinfo {volume} {148}},\ \bibinfo {pages} {175} (\bibinfo {year} {2003})},\ \Eprint {https://arxiv.org/abs/astro-ph/0302209} {arXiv:astro-ph/0302209} \BibitemShut {NoStop}%
\bibitem [{\citenamefont {Aghanim}\ \emph {et~al.}(2020{\natexlab{a}})\citenamefont {Aghanim} \emph {et~al.}}]{collaboration2020planck-5b8}%
  \BibitemOpen
  \bibfield  {author} {\bibinfo {author} {\bibfnamefont {N.}~\bibnamefont {Aghanim}} \emph {et~al.} (\bibinfo {collaboration} {Planck}),\ }\bibfield  {title} {\bibinfo {title} {{Planck 2018 results. I. Overview and the cosmological legacy of Planck}},\ }\href {https://doi.org/10.1051/0004-6361/201833880} {\bibfield  {journal} {\bibinfo  {journal} {Astron. Astrophys.}\ }\textbf {\bibinfo {volume} {641}},\ \bibinfo {pages} {A1} (\bibinfo {year} {2020}{\natexlab{a}})},\ \Eprint {https://arxiv.org/abs/1807.06205} {arXiv:1807.06205 [astro-ph.CO]} \BibitemShut {NoStop}%
\bibitem [{\citenamefont {Akrami}\ \emph {et~al.}(2020)\citenamefont {Akrami} \emph {et~al.}}]{collaboration2020planck-a3c}%
  \BibitemOpen
  \bibfield  {author} {\bibinfo {author} {\bibfnamefont {Y.}~\bibnamefont {Akrami}} \emph {et~al.} (\bibinfo {collaboration} {Planck}),\ }\bibfield  {title} {\bibinfo {title} {{Planck 2018 results. X. Constraints on inflation}},\ }\href {https://doi.org/10.1051/0004-6361/201833887} {\bibfield  {journal} {\bibinfo  {journal} {Astron. Astrophys.}\ }\textbf {\bibinfo {volume} {641}},\ \bibinfo {pages} {A10} (\bibinfo {year} {2020})},\ \Eprint {https://arxiv.org/abs/1807.06211} {arXiv:1807.06211 [astro-ph.CO]} \BibitemShut {NoStop}%
\bibitem [{\citenamefont {Aghanim}\ \emph {et~al.}(2020{\natexlab{b}})\citenamefont {Aghanim} \emph {et~al.}}]{collaboration2019planck-584}%
  \BibitemOpen
  \bibfield  {author} {\bibinfo {author} {\bibfnamefont {N.}~\bibnamefont {Aghanim}} \emph {et~al.} (\bibinfo {collaboration} {Planck}),\ }\bibfield  {title} {\bibinfo {title} {{Planck 2018 results. V. CMB power spectra and likelihoods}},\ }\href {https://doi.org/10.1051/0004-6361/201936386} {\bibfield  {journal} {\bibinfo  {journal} {Astron. Astrophys.}\ }\textbf {\bibinfo {volume} {641}},\ \bibinfo {pages} {A5} (\bibinfo {year} {2020}{\natexlab{b}})},\ \Eprint {https://arxiv.org/abs/1907.12875} {arXiv:1907.12875 [astro-ph.CO]} \BibitemShut {NoStop}%
\bibitem [{\citenamefont {Ade}\ \emph {et~al.}(2021)\citenamefont {Ade} \emph {et~al.}}]{collaboration2021improved-f44}%
  \BibitemOpen
  \bibfield  {author} {\bibinfo {author} {\bibfnamefont {P.~A.~R.}\ \bibnamefont {Ade}} \emph {et~al.} (\bibinfo {collaboration} {BICEP, Keck}),\ }\bibfield  {title} {\bibinfo {title} {{Improved Constraints on Primordial Gravitational Waves using Planck, WMAP, and BICEP/Keck Observations through the 2018 Observing Season}},\ }\href {https://doi.org/10.1103/PhysRevLett.127.151301} {\bibfield  {journal} {\bibinfo  {journal} {Phys. Rev. Lett.}\ }\textbf {\bibinfo {volume} {127}},\ \bibinfo {pages} {151301} (\bibinfo {year} {2021})},\ \Eprint {https://arxiv.org/abs/2110.00483} {arXiv:2110.00483 [astro-ph.CO]} \BibitemShut {NoStop}%
\bibitem [{\citenamefont {Louis}\ \emph {et~al.}(2025)\citenamefont {Louis} \emph {et~al.}}]{louis2025atacama-980}%
  \BibitemOpen
  \bibfield  {author} {\bibinfo {author} {\bibfnamefont {T.}~\bibnamefont {Louis}} \emph {et~al.} (\bibinfo {collaboration} {ACT}),\ }\bibfield  {title} {\bibinfo {title} {{The Atacama Cosmology Telescope: DR6 Power Spectra, Likelihoods and $\Lambda$CDM Parameters}},\ }\Eprint {https://arxiv.org/abs/2503.14452} {arXiv:2503.14452 [astro-ph.CO]} \BibitemShut {NoStop}%
\bibitem [{\citenamefont {Abdul~Karim}\ \emph {et~al.}(2025)\citenamefont {Abdul~Karim} \emph {et~al.}}]{collaboration2025desi-e55}%
  \BibitemOpen
  \bibfield  {author} {\bibinfo {author} {\bibfnamefont {M.}~\bibnamefont {Abdul~Karim}} \emph {et~al.} (\bibinfo {collaboration} {DESI}),\ }\bibfield  {title} {\bibinfo {title} {{DESI DR2 Results II: Measurements of Baryon Acoustic Oscillations and Cosmological Constraints}},\ }\Eprint {https://arxiv.org/abs/2503.14738} {arXiv:2503.14738 [astro-ph.CO]} \BibitemShut {NoStop}%
\bibitem [{\citenamefont {Calabrese}\ \emph {et~al.}(2025)\citenamefont {Calabrese} \emph {et~al.}}]{calabrese2025atacama-a33}%
  \BibitemOpen
  \bibfield  {author} {\bibinfo {author} {\bibfnamefont {E.}~\bibnamefont {Calabrese}} \emph {et~al.} (\bibinfo {collaboration} {ACT}),\ }\bibfield  {title} {\bibinfo {title} {{The Atacama Cosmology Telescope: DR6 Constraints on Extended Cosmological Models}},\ }\Eprint {https://arxiv.org/abs/2503.14454} {arXiv:2503.14454 [astro-ph.CO]} \BibitemShut {NoStop}%
\bibitem [{\citenamefont {Aghanim}\ \emph {et~al.}(2020{\natexlab{c}})\citenamefont {Aghanim}, \citenamefont {Akrami}, \citenamefont {Ashdown},\ and\ \citenamefont {{others}}}]{Planck:2018vyg}%
  \BibitemOpen
  \bibfield  {author} {\bibinfo {author} {\bibfnamefont {N.}~\bibnamefont {Aghanim}}, \bibinfo {author} {\bibfnamefont {Y.}~\bibnamefont {Akrami}}, \bibinfo {author} {\bibfnamefont {M.}~\bibnamefont {Ashdown}},\ and\ \bibinfo {author} {\bibnamefont {{others}}},\ }\bibfield  {title} {\bibinfo {title} {Planck 2018 results. {VI}. {Cosmological} parameters},\ }\href {https://doi.org/10.1051/0004-6361/201833910} {\bibfield  {journal} {\bibinfo  {journal} {Astron. Astrophys.}\ }\textbf {\bibinfo {volume} {641}},\ \bibinfo {pages} {A6} (\bibinfo {year} {2020}{\natexlab{c}})},\ \bibinfo {note} {14633 citations (INSPIRE 2024/8/24) 13585 citations w/o self (INSPIRE 2024/8/24) arXiv:1807.06209 [astro-ph.CO] tex.collaboration: Planck}\BibitemShut {NoStop}%
\bibitem [{\citenamefont {Garcia-Quintero}\ \emph {et~al.}(2025)\citenamefont {Garcia-Quintero} \emph {et~al.}}]{collaboration2025desi-act}%
  \BibitemOpen
  \bibfield  {author} {\bibinfo {author} {\bibfnamefont {C.}~\bibnamefont {Garcia-Quintero}} \emph {et~al.} (\bibinfo {collaboration} {DESI}),\ }\bibfield  {title} {\bibinfo {title} {Cosmological implications of {DESI} {DR}2 {BAO} measurements in light of the latest {ACT} {DR}6 {CMB} data},\ }\Eprint {https://arxiv.org/abs/2504.18464} {arXiv:2504.18464 [astro-ph.CO]} \BibitemShut {NoStop}%
\bibitem [{\citenamefont {Leizerovich}\ \emph {et~al.}(2024)\citenamefont {Leizerovich}, \citenamefont {Landau},\ and\ \citenamefont {Sc\'occola}}]{leizerovich2023tensions-540}%
  \BibitemOpen
  \bibfield  {author} {\bibinfo {author} {\bibfnamefont {M.}~\bibnamefont {Leizerovich}}, \bibinfo {author} {\bibfnamefont {S.~J.}\ \bibnamefont {Landau}},\ and\ \bibinfo {author} {\bibfnamefont {C.~G.}\ \bibnamefont {Sc\'occola}},\ }\bibfield  {title} {\bibinfo {title} {{Tensions in cosmology: A discussion of statistical tools to determine inconsistencies}},\ }\href {https://doi.org/10.1016/j.physletb.2024.138844} {\bibfield  {journal} {\bibinfo  {journal} {Phys. Lett. B}\ }\textbf {\bibinfo {volume} {855}},\ \bibinfo {pages} {138844} (\bibinfo {year} {2024})},\ \Eprint {https://arxiv.org/abs/2312.08542} {arXiv:2312.08542 [astro-ph.CO]} \BibitemShut {NoStop}%
\bibitem [{\citenamefont {Steinhardt}\ \emph {et~al.}(2025)\citenamefont {Steinhardt}, \citenamefont {Phillips},\ and\ \citenamefont {Wojtak}}]{steinhardt2025dark-b8f}%
  \BibitemOpen
  \bibfield  {author} {\bibinfo {author} {\bibfnamefont {C.~L.}\ \bibnamefont {Steinhardt}}, \bibinfo {author} {\bibfnamefont {P.}~\bibnamefont {Phillips}},\ and\ \bibinfo {author} {\bibfnamefont {R.}~\bibnamefont {Wojtak}},\ }\bibfield  {title} {\bibinfo {title} {{Dark Energy Constraints and Joint Cosmological Inference from Mutually Inconsistent Observations}},\ }\Eprint {https://arxiv.org/abs/2504.03829} {arXiv:2504.03829 [astro-ph.CO]} \BibitemShut {NoStop}%
\bibitem [{\citenamefont {Starobinsky}(1980)}]{starobinsky1980new-7c9}%
  \BibitemOpen
  \bibfield  {author} {\bibinfo {author} {\bibfnamefont {A.~A.}\ \bibnamefont {Starobinsky}},\ }\bibfield  {title} {\bibinfo {title} {A new type of isotropic cosmological models without singularity},\ }\href {https://doi.org/10.1016/0370-2693(80)90670-x} {\bibfield  {journal} {\bibinfo  {journal} {Phys. Lett. B}\ }\textbf {\bibinfo {volume} {91}},\ \bibinfo {pages} {99} (\bibinfo {year} {1980})}\BibitemShut {NoStop}%
\bibitem [{\citenamefont {Guth}(1980)}]{guth1980inflationary-4c7}%
  \BibitemOpen
  \bibfield  {author} {\bibinfo {author} {\bibfnamefont {A.~H.}\ \bibnamefont {Guth}},\ }\bibfield  {title} {\bibinfo {title} {{Inflationary universe: A possible solution to the horizon and flatness problems}},\ }\href {https://doi.org/10.1103/physrevd.23.347} {\bibfield  {journal} {\bibinfo  {journal} {Phys. Rev. D}\ }\textbf {\bibinfo {volume} {23}},\ \bibinfo {pages} {347} (\bibinfo {year} {1980})}\BibitemShut {NoStop}%
\bibitem [{\citenamefont {Linde}(1982)}]{Linde:1981mu}%
  \BibitemOpen
  \bibfield  {author} {\bibinfo {author} {\bibfnamefont {A.~D.}\ \bibnamefont {Linde}},\ }\bibfield  {title} {\bibinfo {title} {A new inflationary universe scenario: {{A}} possible solution of the horizon, flatness, momogeneity, isotropy and primordial monopole problems},\ }\href {https://doi.org/10.1016/0370-2693(82)91219-9} {\bibfield  {journal} {\bibinfo  {journal} {Phys. Lett.}\ }\textbf {\bibinfo {volume} {108B}},\ \bibinfo {pages} {389} (\bibinfo {year} {1982})}\BibitemShut {NoStop}%
\bibitem [{\citenamefont {Starobinsky}(1982)}]{Starobinsky:1982ee}%
  \BibitemOpen
  \bibfield  {author} {\bibinfo {author} {\bibfnamefont {A.~A.}\ \bibnamefont {Starobinsky}},\ }\bibfield  {title} {\bibinfo {title} {Dynamics of phase transition in the new inflationary universe scenario and generation of perturbations},\ }\href {https://doi.org/10.1016/0370-2693(82)90541-X} {\bibfield  {journal} {\bibinfo  {journal} {Phys. Lett.}\ }\textbf {\bibinfo {volume} {117B}},\ \bibinfo {pages} {175} (\bibinfo {year} {1982})}\BibitemShut {NoStop}%
\bibitem [{\citenamefont {Albrecht}\ and\ \citenamefont {Steinhardt}(1982)}]{Albrecht:1982wi}%
  \BibitemOpen
  \bibfield  {author} {\bibinfo {author} {\bibfnamefont {A.}~\bibnamefont {Albrecht}}\ and\ \bibinfo {author} {\bibfnamefont {P.~J.}\ \bibnamefont {Steinhardt}},\ }\bibfield  {title} {\bibinfo {title} {{Cosmology} for {Grand} {Unified} {Theories} with {Radiatively} {Induced} {Symmetry} {Breaking}},\ }\href {https://doi.org/10.1103/PhysRevLett.48.1220} {\bibfield  {journal} {\bibinfo  {journal} {Phys. Rev. Lett.}\ }\textbf {\bibinfo {volume} {48}},\ \bibinfo {pages} {1220} (\bibinfo {year} {1982})}\BibitemShut {NoStop}%
\bibitem [{\citenamefont {Linde}(1983)}]{Linde:1983gd}%
  \BibitemOpen
  \bibfield  {author} {\bibinfo {author} {\bibfnamefont {A.~D.}\ \bibnamefont {Linde}},\ }\bibfield  {title} {\bibinfo {title} {Chaotic inflation},\ }\href {https://doi.org/10.1016/0370-2693(83)90837-7} {\bibfield  {journal} {\bibinfo  {journal} {Phys. Lett.}\ }\textbf {\bibinfo {volume} {129B}},\ \bibinfo {pages} {177} (\bibinfo {year} {1983})}\BibitemShut {NoStop}%
\bibitem [{\citenamefont {Starobinsky}(1979)}]{Starobinsky:1979ty}%
  \BibitemOpen
  \bibfield  {author} {\bibinfo {author} {\bibfnamefont {A.~A.}\ \bibnamefont {Starobinsky}},\ }\bibfield  {title} {\bibinfo {title} {Spectrum of relict gravitational radiation and the early state of the universe},\ }\href@noop {} {\bibfield  {journal} {\bibinfo  {journal} {JETP Lett.}\ }\textbf {\bibinfo {volume} {30}},\ \bibinfo {pages} {682} (\bibinfo {year} {1979})}\BibitemShut {NoStop}%
\bibitem [{\citenamefont {Mukhanov}\ and\ \citenamefont {Chibisov}(1981)}]{Mukhanov:1981xt}%
  \BibitemOpen
  \bibfield  {author} {\bibinfo {author} {\bibfnamefont {V.~F.}\ \bibnamefont {Mukhanov}}\ and\ \bibinfo {author} {\bibfnamefont {G.~V.}\ \bibnamefont {Chibisov}},\ }\bibfield  {title} {\bibinfo {title} {Quantum fluctuations and a nonsingular universe},\ }\href@noop {} {\bibfield  {journal} {\bibinfo  {journal} {JETP Lett.}\ }\textbf {\bibinfo {volume} {33}},\ \bibinfo {pages} {532} (\bibinfo {year} {1981})}\BibitemShut {NoStop}%
\bibitem [{\citenamefont {Mukhanov}\ and\ \citenamefont {Chibisov}(1982)}]{Mukhanov:1982nu}%
  \BibitemOpen
  \bibfield  {author} {\bibinfo {author} {\bibfnamefont {V.~F.}\ \bibnamefont {Mukhanov}}\ and\ \bibinfo {author} {\bibfnamefont {G.~V.}\ \bibnamefont {Chibisov}},\ }\bibfield  {title} {\bibinfo {title} {Vacuum energy and large scale structure of the {Universe}},\ }\href@noop {} {\bibfield  {journal} {\bibinfo  {journal} {Sov. Phys. JETP}\ }\textbf {\bibinfo {volume} {56}},\ \bibinfo {pages} {258} (\bibinfo {year} {1982})}\BibitemShut {NoStop}%
\bibitem [{\citenamefont {Guth}\ and\ \citenamefont {Pi}(1982)}]{Guth:1982ec}%
  \BibitemOpen
  \bibfield  {author} {\bibinfo {author} {\bibfnamefont {A.~H.}\ \bibnamefont {Guth}}\ and\ \bibinfo {author} {\bibfnamefont {S.~Y.}\ \bibnamefont {Pi}},\ }\bibfield  {title} {\bibinfo {title} {Fluctuations in the {New} {Inflationary} {Universe}},\ }\href {https://doi.org/10.1103/PhysRevLett.49.1110} {\bibfield  {journal} {\bibinfo  {journal} {Phys. Rev. Lett.}\ }\textbf {\bibinfo {volume} {49}},\ \bibinfo {pages} {1110} (\bibinfo {year} {1982})}\BibitemShut {NoStop}%
\bibitem [{\citenamefont {Hawking}(1982)}]{Hawking:1982cz}%
  \BibitemOpen
  \bibfield  {author} {\bibinfo {author} {\bibfnamefont {S.~W.}\ \bibnamefont {Hawking}},\ }\bibfield  {title} {\bibinfo {title} {The development of irregularities in a single bubble inflationary universe},\ }\href {https://doi.org/10.1016/0370-2693(82)90373-2} {\bibfield  {journal} {\bibinfo  {journal} {Phys. Lett.}\ }\textbf {\bibinfo {volume} {115B}},\ \bibinfo {pages} {295} (\bibinfo {year} {1982})}\BibitemShut {NoStop}%
\bibitem [{\citenamefont {Bardeen}\ \emph {et~al.}(1983)\citenamefont {Bardeen}, \citenamefont {Steinhardt},\ and\ \citenamefont {Turner}}]{Bardeen:1983qw}%
  \BibitemOpen
  \bibfield  {author} {\bibinfo {author} {\bibfnamefont {J.~M.}\ \bibnamefont {Bardeen}}, \bibinfo {author} {\bibfnamefont {P.~J.}\ \bibnamefont {Steinhardt}},\ and\ \bibinfo {author} {\bibfnamefont {M.~S.}\ \bibnamefont {Turner}},\ }\bibfield  {title} {\bibinfo {title} {Spontaneous creation of almost scale-free density perturbations in an inflationary universe},\ }\href {https://doi.org/10.1103/PhysRevD.28.679} {\bibfield  {journal} {\bibinfo  {journal} {Phys. Rev. D}\ }\textbf {\bibinfo {volume} {28}},\ \bibinfo {pages} {679} (\bibinfo {year} {1983})}\BibitemShut {NoStop}%
\bibitem [{\citenamefont {Martin}\ \emph {et~al.}(2014)\citenamefont {Martin}, \citenamefont {Ringeval},\ and\ \citenamefont {Vennin}}]{martin2013encyclopaedia-a01}%
  \BibitemOpen
  \bibfield  {author} {\bibinfo {author} {\bibfnamefont {J.}~\bibnamefont {Martin}}, \bibinfo {author} {\bibfnamefont {C.}~\bibnamefont {Ringeval}},\ and\ \bibinfo {author} {\bibfnamefont {V.}~\bibnamefont {Vennin}},\ }\bibfield  {title} {\bibinfo {title} {{Encyclop\ae{}dia Inflationaris}: {Opiparous Edition}},\ }\href {https://doi.org/10.1016/j.dark.2024.101653} {\bibfield  {journal} {\bibinfo  {journal} {Phys. Dark Univ.}\ }\textbf {\bibinfo {volume} {5-6}},\ \bibinfo {pages} {75} (\bibinfo {year} {2014})},\ \Eprint {https://arxiv.org/abs/1303.3787} {arXiv:1303.3787 [astro-ph.CO]} \BibitemShut {NoStop}%
\bibitem [{\citenamefont {Bezrukov}\ and\ \citenamefont {Shaposhnikov}(2008)}]{bezrukov2008standard-957}%
  \BibitemOpen
  \bibfield  {author} {\bibinfo {author} {\bibfnamefont {F.}~\bibnamefont {Bezrukov}}\ and\ \bibinfo {author} {\bibfnamefont {M.}~\bibnamefont {Shaposhnikov}},\ }\bibfield  {title} {\bibinfo {title} {{The Standard Model Higgs boson as the inflaton}},\ }\href {https://doi.org/10.1016/j.physletb.2007.11.072} {\bibfield  {journal} {\bibinfo  {journal} {Phys. Lett. B}\ }\textbf {\bibinfo {volume} {659}},\ \bibinfo {pages} {703} (\bibinfo {year} {2008})},\ \Eprint {https://arxiv.org/abs/0710.3755} {0710.3755} \BibitemShut {NoStop}%
\bibitem [{\citenamefont {Martin}\ \emph {et~al.}(2024)\citenamefont {Martin}, \citenamefont {Ringeval},\ and\ \citenamefont {Vennin}}]{martin2024cosmic-687}%
  \BibitemOpen
  \bibfield  {author} {\bibinfo {author} {\bibfnamefont {J.}~\bibnamefont {Martin}}, \bibinfo {author} {\bibfnamefont {C.}~\bibnamefont {Ringeval}},\ and\ \bibinfo {author} {\bibfnamefont {V.}~\bibnamefont {Vennin}},\ }\bibfield  {title} {\bibinfo {title} {{Cosmic Inflation at the crossroads}},\ }\href {https://doi.org/10.1088/1475-7516/2024/07/087} {\bibfield  {journal} {\bibinfo  {journal} {J. Cosmol. Astropart. Phys.}\ }\bibinfo {volume} {07} (\bibinfo {year} {2024})\ \bibinfo {pages} {087}},\ \Eprint {https://arxiv.org/abs/2404.10647} {arXiv:2404.10647 [astro-ph.CO]} \BibitemShut {NoStop}%
\bibitem [{\citenamefont {Chakraborty}\ \emph {et~al.}(2023)\citenamefont {Chakraborty}, \citenamefont {Haque}, \citenamefont {Maity},\ and\ \citenamefont {Mondal}}]{Chakraborty:2023ocr}%
  \BibitemOpen
  \bibfield  {author} {\bibinfo {author} {\bibfnamefont {A.}~\bibnamefont {Chakraborty}}, \bibinfo {author} {\bibfnamefont {M.~R.}\ \bibnamefont {Haque}}, \bibinfo {author} {\bibfnamefont {D.}~\bibnamefont {Maity}},\ and\ \bibinfo {author} {\bibfnamefont {R.}~\bibnamefont {Mondal}},\ }\bibfield  {title} {\bibinfo {title} {{Inflaton phenomenology via reheating in light of primordial gravitational waves and the latest BICEP/Keck data}},\ }\href {https://doi.org/10.1103/PhysRevD.108.023515} {\bibfield  {journal} {\bibinfo  {journal} {Phys. Rev. D}\ }\textbf {\bibinfo {volume} {108}},\ \bibinfo {pages} {023515} (\bibinfo {year} {2023})},\ \Eprint {https://arxiv.org/abs/2304.13637} {arXiv:2304.13637 [astro-ph.CO]} \BibitemShut {NoStop}%
\bibitem [{\citenamefont {Kofman}\ \emph {et~al.}(1997)\citenamefont {Kofman}, \citenamefont {Linde},\ and\ \citenamefont {Starobinsky}}]{Kofman:1997yn}%
  \BibitemOpen
  \bibfield  {author} {\bibinfo {author} {\bibfnamefont {L.}~\bibnamefont {Kofman}}, \bibinfo {author} {\bibfnamefont {A.~D.}\ \bibnamefont {Linde}},\ and\ \bibinfo {author} {\bibfnamefont {A.~A.}\ \bibnamefont {Starobinsky}},\ }\bibfield  {title} {\bibinfo {title} {{Towards the theory of reheating after inflation}},\ }\href {https://doi.org/10.1103/PhysRevD.56.3258} {\bibfield  {journal} {\bibinfo  {journal} {Phys. Rev. D}\ }\textbf {\bibinfo {volume} {56}},\ \bibinfo {pages} {3258} (\bibinfo {year} {1997})},\ \Eprint {https://arxiv.org/abs/hep-ph/9704452} {arXiv:hep-ph/9704452} \BibitemShut {NoStop}%
\bibitem [{\citenamefont {Allahverdi}\ \emph {et~al.}(2010)\citenamefont {Allahverdi}, \citenamefont {Brandenberger}, \citenamefont {Cyr-Racine},\ and\ \citenamefont {Mazumdar}}]{Allahverdi:2010xz}%
  \BibitemOpen
  \bibfield  {author} {\bibinfo {author} {\bibfnamefont {R.}~\bibnamefont {Allahverdi}}, \bibinfo {author} {\bibfnamefont {R.}~\bibnamefont {Brandenberger}}, \bibinfo {author} {\bibfnamefont {F.-Y.}\ \bibnamefont {Cyr-Racine}},\ and\ \bibinfo {author} {\bibfnamefont {A.}~\bibnamefont {Mazumdar}},\ }\bibfield  {title} {\bibinfo {title} {{Reheating in Inflationary Cosmology: Theory and Applications}},\ }\href {https://doi.org/10.1146/annurev.nucl.012809.104511} {\bibfield  {journal} {\bibinfo  {journal} {Ann. Rev. Nucl. Part. Sci.}\ }\textbf {\bibinfo {volume} {60}},\ \bibinfo {pages} {27} (\bibinfo {year} {2010})},\ \Eprint {https://arxiv.org/abs/1001.2600} {arXiv:1001.2600 [hep-th]} \BibitemShut {NoStop}%
\bibitem [{\citenamefont {Amin}\ \emph {et~al.}(2014)\citenamefont {Amin}, \citenamefont {Hertzberg}, \citenamefont {Kaiser},\ and\ \citenamefont {Karouby}}]{Amin:2014eta}%
  \BibitemOpen
  \bibfield  {author} {\bibinfo {author} {\bibfnamefont {M.~A.}\ \bibnamefont {Amin}}, \bibinfo {author} {\bibfnamefont {M.~P.}\ \bibnamefont {Hertzberg}}, \bibinfo {author} {\bibfnamefont {D.~I.}\ \bibnamefont {Kaiser}},\ and\ \bibinfo {author} {\bibfnamefont {J.}~\bibnamefont {Karouby}},\ }\bibfield  {title} {\bibinfo {title} {{Nonperturbative Dynamics Of Reheating After Inflation: A Review}},\ }\href {https://doi.org/10.1142/S0218271815300037} {\bibfield  {journal} {\bibinfo  {journal} {Int. J. Mod. Phys. D}\ }\textbf {\bibinfo {volume} {24}},\ \bibinfo {pages} {1530003} (\bibinfo {year} {2014})},\ \Eprint {https://arxiv.org/abs/1410.3808} {arXiv:1410.3808 [hep-ph]} \BibitemShut {NoStop}%
\bibitem [{\citenamefont {Ellis}\ \emph {et~al.}(2022)\citenamefont {Ellis}, \citenamefont {Garcia}, \citenamefont {Nanopoulos}, \citenamefont {Olive},\ and\ \citenamefont {Verner}}]{ellis2022bicepkeck-6ed}%
  \BibitemOpen
  \bibfield  {author} {\bibinfo {author} {\bibfnamefont {J.}~\bibnamefont {Ellis}}, \bibinfo {author} {\bibfnamefont {M.~A.~G.}\ \bibnamefont {Garcia}}, \bibinfo {author} {\bibfnamefont {D.~V.}\ \bibnamefont {Nanopoulos}}, \bibinfo {author} {\bibfnamefont {K.~A.}\ \bibnamefont {Olive}},\ and\ \bibinfo {author} {\bibfnamefont {S.}~\bibnamefont {Verner}},\ }\bibfield  {title} {\bibinfo {title} {{BICEP}/keck constraints on attractor models of inflation and reheating},\ }\href {https://doi.org/10.1103/physrevd.105.043504} {\bibfield  {journal} {\bibinfo  {journal} {Phys. Rev. D}\ }\textbf {\bibinfo {volume} {105}},\ \bibinfo {pages} {043504} (\bibinfo {year} {2022})},\ \Eprint {https://arxiv.org/abs/2112.04466} {2112.04466} \BibitemShut {NoStop}%
\bibitem [{\citenamefont {Blas}\ \emph {et~al.}(2011)\citenamefont {Blas}, \citenamefont {Lesgourgues},\ and\ \citenamefont {Tram}}]{blas2011cosmic-8f1}%
  \BibitemOpen
  \bibfield  {author} {\bibinfo {author} {\bibfnamefont {D.}~\bibnamefont {Blas}}, \bibinfo {author} {\bibfnamefont {J.}~\bibnamefont {Lesgourgues}},\ and\ \bibinfo {author} {\bibfnamefont {T.}~\bibnamefont {Tram}},\ }\bibfield  {title} {\bibinfo {title} {{The Cosmic Linear Anisotropy Solving System (CLASS) II: Approximation schemes}},\ }\href {https://doi.org/10.1088/1475-7516/2011/07/034} {\bibfield  {journal} {\bibinfo  {journal} {J. Cosmol. Astropart. Phys.}\ }\bibinfo {volume} {07} (\bibinfo {year} {2011})\ \bibinfo {pages} {034}},\ \Eprint {https://arxiv.org/abs/1104.2933} {arXiv:1104.2933 [astro-ph.CO]} \BibitemShut {NoStop}%
\bibitem [{\citenamefont {Lesgourgues}(2011)}]{lesgourgues2011cosmic-c31}%
  \BibitemOpen
  \bibfield  {author} {\bibinfo {author} {\bibfnamefont {J.}~\bibnamefont {Lesgourgues}},\ }\bibfield  {title} {\bibinfo {title} {The cosmic linear anisotropy solving system ({CLASS}) i: Overview},\ }\Eprint {https://arxiv.org/abs/1104.2932} {1104.2932} \BibitemShut {NoStop}%
\bibitem [{\citenamefont {Torrado}\ and\ \citenamefont {Lewis}(2021)}]{torrado2020cobaya-7d4}%
  \BibitemOpen
  \bibfield  {author} {\bibinfo {author} {\bibfnamefont {J.}~\bibnamefont {Torrado}}\ and\ \bibinfo {author} {\bibfnamefont {A.}~\bibnamefont {Lewis}},\ }\bibfield  {title} {\bibinfo {title} {{Cobaya: Code for Bayesian Analysis of hierarchical physical models}},\ }\href {https://doi.org/10.1088/1475-7516/2021/05/057} {\bibfield  {journal} {\bibinfo  {journal} {J. Cosmol. Astropart. Phys.}\ }\bibinfo {volume} {05} (\bibinfo {year} {2021})\ \bibinfo {pages} {057}},\ \Eprint {https://arxiv.org/abs/2005.05290} {arXiv:2005.05290 [astro-ph.IM]} \BibitemShut {NoStop}%
\bibitem [{\citenamefont {Lewis}(2019)}]{lewis2019getdist-ed7}%
  \BibitemOpen
  \bibfield  {author} {\bibinfo {author} {\bibfnamefont {A.}~\bibnamefont {Lewis}},\ }\bibfield  {title} {\bibinfo {title} {{GetDist}: a python package for analysing monte carlo samples},\ }\Eprint {https://arxiv.org/abs/1910.13970} {1910.13970} \BibitemShut {NoStop}%
\bibitem [{\citenamefont {Adame}\ \emph {et~al.}(2025{\natexlab{a}})\citenamefont {Adame} \emph {et~al.}}]{collaboration2024desi-2ba}%
  \BibitemOpen
  \bibfield  {author} {\bibinfo {author} {\bibfnamefont {A.~G.}\ \bibnamefont {Adame}} \emph {et~al.} (\bibinfo {collaboration} {DESI}),\ }\bibfield  {title} {\bibinfo {title} {{DESI 2024 III: Baryon Acoustic Oscillations from galaxies and quasars}},\ }\href {https://doi.org/10.1088/1475-7516/2025/04/012} {\bibfield  {journal} {\bibinfo  {journal} {J. Cosmol. Astropart. Phys.}\ }\bibinfo {volume} {04} (\bibinfo {year} {2025}{\natexlab{a}})\ \bibinfo {pages} {012}},\ \Eprint {https://arxiv.org/abs/2404.03000} {arXiv:2404.03000 [astro-ph.CO]} \BibitemShut {NoStop}%
\bibitem [{\citenamefont {Adame}\ \emph {et~al.}(2025{\natexlab{b}})\citenamefont {Adame} \emph {et~al.}}]{collaboration2024desi-020}%
  \BibitemOpen
  \bibfield  {author} {\bibinfo {author} {\bibfnamefont {A.~G.}\ \bibnamefont {Adame}} \emph {et~al.} (\bibinfo {collaboration} {DESI}),\ }\bibfield  {title} {\bibinfo {title} {{DESI 2024 IV: Baryon Acoustic Oscillations from the Lyman alpha forest}},\ }\href {https://doi.org/10.1088/1475-7516/2025/01/124} {\bibfield  {journal} {\bibinfo  {journal} {J. Cosmol. Astropart. Phys.}\ }\bibinfo {volume} {01} (\bibinfo {year} {2025}{\natexlab{b}})\ \bibinfo {pages} {124}},\ \Eprint {https://arxiv.org/abs/2404.03001} {arXiv:2404.03001 [astro-ph.CO]} \BibitemShut {NoStop}%
\bibitem [{\citenamefont {Adame}\ \emph {et~al.}(2025{\natexlab{c}})\citenamefont {Adame} \emph {et~al.}}]{collaboration2024desi-265}%
  \BibitemOpen
  \bibfield  {author} {\bibinfo {author} {\bibfnamefont {A.~G.}\ \bibnamefont {Adame}} \emph {et~al.} (\bibinfo {collaboration} {DESI}),\ }\bibfield  {title} {\bibinfo {title} {{DESI 2024 VI: cosmological constraints from the measurements of baryon acoustic oscillations}},\ }\href {https://doi.org/10.1088/1475-7516/2025/02/021} {\bibfield  {journal} {\bibinfo  {journal} {J. Cosmol. Astropart. Phys.}\ }\bibinfo {volume} {02} (\bibinfo {year} {2025}{\natexlab{c}})\ \bibinfo {pages} {021}},\ \Eprint {https://arxiv.org/abs/2404.03002} {arXiv:2404.03002 [astro-ph.CO]} \BibitemShut {NoStop}%
\bibitem [{\citenamefont {Carron}\ \emph {et~al.}(2022)\citenamefont {Carron}, \citenamefont {Mirmelstein},\ and\ \citenamefont {Lewis}}]{carron2022cmb-ee4}%
  \BibitemOpen
  \bibfield  {author} {\bibinfo {author} {\bibfnamefont {J.}~\bibnamefont {Carron}}, \bibinfo {author} {\bibfnamefont {M.}~\bibnamefont {Mirmelstein}},\ and\ \bibinfo {author} {\bibfnamefont {A.}~\bibnamefont {Lewis}},\ }\bibfield  {title} {\bibinfo {title} {{CMB lensing from Planck PR4~maps}},\ }\href {https://doi.org/10.1088/1475-7516/2022/09/039} {\bibfield  {journal} {\bibinfo  {journal} {J. Cosmol. Astropart. Phys.}\ }\bibinfo {volume} {09} (\bibinfo {year} {2022})\ \bibinfo {pages} {039}},\ \Eprint {https://arxiv.org/abs/2206.07773} {arXiv:2206.07773 [astro-ph.CO]} \BibitemShut {NoStop}%
\bibitem [{\citenamefont {Kullback}\ and\ \citenamefont {Leibler}(1951)}]{kullback1951information-d27}%
  \BibitemOpen
  \bibfield  {author} {\bibinfo {author} {\bibfnamefont {S.}~\bibnamefont {Kullback}}\ and\ \bibinfo {author} {\bibfnamefont {R.~A.}\ \bibnamefont {Leibler}},\ }\bibfield  {title} {\bibinfo {title} {On information and sufficiency},\ }\href {https://doi.org/10.1214/aoms/1177729694} {\bibfield  {journal} {\bibinfo  {journal} {Ann. Math. Stat.}\ }\textbf {\bibinfo {volume} {22}},\ \bibinfo {pages} {79} (\bibinfo {year} {1951})}\BibitemShut {NoStop}%
\bibitem [{\citenamefont {Virtanen}\ \emph {et~al.}(2020)\citenamefont {Virtanen}, \citenamefont {Gommers}, \citenamefont {Oliphant},\ and\ \citenamefont {{others}}}]{SciPy2020}%
  \BibitemOpen
  \bibfield  {author} {\bibinfo {author} {\bibfnamefont {P.}~\bibnamefont {Virtanen}}, \bibinfo {author} {\bibfnamefont {R.}~\bibnamefont {Gommers}}, \bibinfo {author} {\bibfnamefont {T.~E.}\ \bibnamefont {Oliphant}},\ and\ \bibinfo {author} {\bibnamefont {{others}}},\ }\bibfield  {title} {\bibinfo {title} {{{SciPy} 1.0: Fundamental Algorithms for Scientific Computing in Python}},\ }\href {https://doi.org/10.1038/s41592-019-0686-2} {\bibfield  {journal} {\bibinfo  {journal} {Nature Methods}\ }\textbf {\bibinfo {volume} {17}},\ \bibinfo {pages} {261} (\bibinfo {year} {2020})}\BibitemShut {NoStop}%
\bibitem [{\citenamefont {Gorbunov}\ and\ \citenamefont {Panin}(2011)}]{Gorbunov:2010bn}%
  \BibitemOpen
  \bibfield  {author} {\bibinfo {author} {\bibfnamefont {D.~S.}\ \bibnamefont {Gorbunov}}\ and\ \bibinfo {author} {\bibfnamefont {A.~G.}\ \bibnamefont {Panin}},\ }\bibfield  {title} {\bibinfo {title} {{Scalaron the mighty: producing dark matter and baryon asymmetry at reheating}},\ }\href {https://doi.org/10.1016/j.physletb.2011.04.067} {\bibfield  {journal} {\bibinfo  {journal} {Phys. Lett. B}\ }\textbf {\bibinfo {volume} {700}},\ \bibinfo {pages} {157} (\bibinfo {year} {2011})},\ \Eprint {https://arxiv.org/abs/1009.2448} {arXiv:1009.2448 [hep-ph]} \BibitemShut {NoStop}%
\bibitem [{\citenamefont {Jeong}\ \emph {et~al.}(2023)\citenamefont {Jeong}, \citenamefont {Kamada}, \citenamefont {Starobinsky},\ and\ \citenamefont {Yokoyama}}]{Jeong:2023zrv}%
  \BibitemOpen
  \bibfield  {author} {\bibinfo {author} {\bibfnamefont {H.}~\bibnamefont {Jeong}}, \bibinfo {author} {\bibfnamefont {K.}~\bibnamefont {Kamada}}, \bibinfo {author} {\bibfnamefont {A.~A.}\ \bibnamefont {Starobinsky}},\ and\ \bibinfo {author} {\bibfnamefont {J.}~\bibnamefont {Yokoyama}},\ }\bibfield  {title} {\bibinfo {title} {{Reheating process in the R $^{2}$ inflationary model with the baryogenesis scenario}},\ }\href {https://doi.org/10.1088/1475-7516/2023/11/023} {\bibfield  {journal} {\bibinfo  {journal} {JCAP}\ }\bibinfo {volume} {11} (\bibinfo {year} {2023})\ \bibinfo {pages} {023}},\ \Eprint {https://arxiv.org/abs/2305.14273} {arXiv:2305.14273 [hep-ph]} \BibitemShut {NoStop}%
\bibitem [{\citenamefont {Bezrukov}\ \emph {et~al.}(2009)\citenamefont {Bezrukov}, \citenamefont {Gorbunov},\ and\ \citenamefont {Shaposhnikov}}]{Bezrukov:2008ut}%
  \BibitemOpen
  \bibfield  {author} {\bibinfo {author} {\bibfnamefont {F.}~\bibnamefont {Bezrukov}}, \bibinfo {author} {\bibfnamefont {D.}~\bibnamefont {Gorbunov}},\ and\ \bibinfo {author} {\bibfnamefont {M.}~\bibnamefont {Shaposhnikov}},\ }\bibfield  {title} {\bibinfo {title} {{On initial conditions for the Hot Big Bang}},\ }\href {https://doi.org/10.1088/1475-7516/2009/06/029} {\bibfield  {journal} {\bibinfo  {journal} {JCAP}\ }\bibinfo {volume} {06} (\bibinfo {year} {2009})\ \bibinfo {pages} {029}},\ \Eprint {https://arxiv.org/abs/0812.3622} {arXiv:0812.3622 [hep-ph]} \BibitemShut {NoStop}%
\bibitem [{\citenamefont {Garcia-Bellido}\ \emph {et~al.}(2009)\citenamefont {Garcia-Bellido}, \citenamefont {Figueroa},\ and\ \citenamefont {Rubio}}]{Garcia-Bellido:2008ycs}%
  \BibitemOpen
  \bibfield  {author} {\bibinfo {author} {\bibfnamefont {J.}~\bibnamefont {Garcia-Bellido}}, \bibinfo {author} {\bibfnamefont {D.~G.}\ \bibnamefont {Figueroa}},\ and\ \bibinfo {author} {\bibfnamefont {J.}~\bibnamefont {Rubio}},\ }\bibfield  {title} {\bibinfo {title} {{Preheating in the Standard Model with the Higgs-Inflaton coupled to gravity}},\ }\href {https://doi.org/10.1103/PhysRevD.79.063531} {\bibfield  {journal} {\bibinfo  {journal} {Phys. Rev. D}\ }\textbf {\bibinfo {volume} {79}},\ \bibinfo {pages} {063531} (\bibinfo {year} {2009})},\ \Eprint {https://arxiv.org/abs/0812.4624} {arXiv:0812.4624 [hep-ph]} \BibitemShut {NoStop}%
\bibitem [{\citenamefont {Bezrukov}\ \emph {et~al.}(2011)\citenamefont {Bezrukov}, \citenamefont {Gorbunov},\ and\ \citenamefont {Shaposhnikov}}]{Bezrukov:2011sz}%
  \BibitemOpen
  \bibfield  {author} {\bibinfo {author} {\bibfnamefont {F.}~\bibnamefont {Bezrukov}}, \bibinfo {author} {\bibfnamefont {D.}~\bibnamefont {Gorbunov}},\ and\ \bibinfo {author} {\bibfnamefont {M.}~\bibnamefont {Shaposhnikov}},\ }\bibfield  {title} {\bibinfo {title} {{Late and early time phenomenology of Higgs-dependent cutoff}},\ }\href {https://doi.org/10.1088/1475-7516/2011/10/001} {\bibfield  {journal} {\bibinfo  {journal} {JCAP}\ }\bibinfo {volume} {10} (\bibinfo {year} {2011})\ \bibinfo {pages} {001}},\ \Eprint {https://arxiv.org/abs/1106.5019} {arXiv:1106.5019 [hep-ph]} \BibitemShut {NoStop}%
\bibitem [{\citenamefont {Ema}\ \emph {et~al.}(2017)\citenamefont {Ema}, \citenamefont {Jinno}, \citenamefont {Mukaida},\ and\ \citenamefont {Nakayama}}]{Ema:2016dny}%
  \BibitemOpen
  \bibfield  {author} {\bibinfo {author} {\bibfnamefont {Y.}~\bibnamefont {Ema}}, \bibinfo {author} {\bibfnamefont {R.}~\bibnamefont {Jinno}}, \bibinfo {author} {\bibfnamefont {K.}~\bibnamefont {Mukaida}},\ and\ \bibinfo {author} {\bibfnamefont {K.}~\bibnamefont {Nakayama}},\ }\bibfield  {title} {\bibinfo {title} {{Violent Preheating in Inflation with Nonminimal Coupling}},\ }\href {https://doi.org/10.1088/1475-7516/2017/02/045} {\bibfield  {journal} {\bibinfo  {journal} {JCAP}\ }\bibinfo {volume} {02} (\bibinfo {year} {2017})\ \bibinfo {pages} {045}},\ \Eprint {https://arxiv.org/abs/1609.05209} {arXiv:1609.05209 [hep-ph]} \BibitemShut {NoStop}%
\bibitem [{\citenamefont {DeCross}\ \emph {et~al.}(2018)\citenamefont {DeCross}, \citenamefont {Kaiser}, \citenamefont {Prabhu}, \citenamefont {Prescod-Weinstein},\ and\ \citenamefont {Sfakianakis}}]{DeCross:2016cbs}%
  \BibitemOpen
  \bibfield  {author} {\bibinfo {author} {\bibfnamefont {M.~P.}\ \bibnamefont {DeCross}}, \bibinfo {author} {\bibfnamefont {D.~I.}\ \bibnamefont {Kaiser}}, \bibinfo {author} {\bibfnamefont {A.}~\bibnamefont {Prabhu}}, \bibinfo {author} {\bibfnamefont {C.}~\bibnamefont {Prescod-Weinstein}},\ and\ \bibinfo {author} {\bibfnamefont {E.~I.}\ \bibnamefont {Sfakianakis}},\ }\bibfield  {title} {\bibinfo {title} {{Preheating after multifield inflation with nonminimal couplings, III: Dynamical spacetime results}},\ }\href {https://doi.org/10.1103/PhysRevD.97.023528} {\bibfield  {journal} {\bibinfo  {journal} {Phys. Rev. D}\ }\textbf {\bibinfo {volume} {97}},\ \bibinfo {pages} {023528} (\bibinfo {year} {2018})},\ \Eprint {https://arxiv.org/abs/1610.08916} {arXiv:1610.08916 [astro-ph.CO]} \BibitemShut {NoStop}%
\bibitem [{\citenamefont {Turner}(1983)}]{Turner:1983he}%
  \BibitemOpen
  \bibfield  {author} {\bibinfo {author} {\bibfnamefont {M.~S.}\ \bibnamefont {Turner}},\ }\bibfield  {title} {\bibinfo {title} {{Coherent Scalar Field Oscillations in an Expanding Universe}},\ }\href {https://doi.org/10.1103/PhysRevD.28.1243} {\bibfield  {journal} {\bibinfo  {journal} {Phys. Rev. D}\ }\textbf {\bibinfo {volume} {28}},\ \bibinfo {pages} {1243} (\bibinfo {year} {1983})}\BibitemShut {NoStop}%
\bibitem [{\citenamefont {Joyce}(1997)}]{Joyce:1996cp}%
  \BibitemOpen
  \bibfield  {author} {\bibinfo {author} {\bibfnamefont {M.}~\bibnamefont {Joyce}},\ }\bibfield  {title} {\bibinfo {title} {{Electroweak Baryogenesis and the Expansion Rate of the Universe}},\ }\href {https://doi.org/10.1103/PhysRevD.55.1875} {\bibfield  {journal} {\bibinfo  {journal} {Phys. Rev. D}\ }\textbf {\bibinfo {volume} {55}},\ \bibinfo {pages} {1875} (\bibinfo {year} {1997})},\ \Eprint {https://arxiv.org/abs/hep-ph/9606223} {arXiv:hep-ph/9606223} \BibitemShut {NoStop}%
\bibitem [{\citenamefont {Drees}\ and\ \citenamefont {Xu}(2025)}]{Drees:2025ngb}%
  \BibitemOpen
  \bibfield  {author} {\bibinfo {author} {\bibfnamefont {M.}~\bibnamefont {Drees}}\ and\ \bibinfo {author} {\bibfnamefont {Y.}~\bibnamefont {Xu}},\ }\bibfield  {title} {\bibinfo {title} {{Refined Predictions for Starobinsky Inflation and Post-inflationary Constraints in Light of ACT}},\ }\Eprint {https://arxiv.org/abs/2504.20757} {arXiv:2504.20757 [astro-ph.CO]} \BibitemShut {NoStop}%
\bibitem [{\citenamefont {Liu}\ \emph {et~al.}(2025)\citenamefont {Liu}, \citenamefont {Yi},\ and\ \citenamefont {Gong}}]{Liu:2025qca}%
  \BibitemOpen
  \bibfield  {author} {\bibinfo {author} {\bibfnamefont {L.}~\bibnamefont {Liu}}, \bibinfo {author} {\bibfnamefont {Z.}~\bibnamefont {Yi}},\ and\ \bibinfo {author} {\bibfnamefont {Y.}~\bibnamefont {Gong}},\ }\bibfield  {title} {\bibinfo {title} {{Reconciling Higgs Inflation with ACT Observations through Reheating}},\ }\Eprint {https://arxiv.org/abs/2505.02407} {arXiv:2505.02407 [astro-ph.CO]} \BibitemShut {NoStop}%
\bibitem [{\citenamefont {Haque}\ \emph {et~al.}(2025)\citenamefont {Haque}, \citenamefont {Pal},\ and\ \citenamefont {Paul}}]{Haque:2025uis}%
  \BibitemOpen
  \bibfield  {author} {\bibinfo {author} {\bibfnamefont {M.~R.}\ \bibnamefont {Haque}}, \bibinfo {author} {\bibfnamefont {S.}~\bibnamefont {Pal}},\ and\ \bibinfo {author} {\bibfnamefont {D.}~\bibnamefont {Paul}},\ }\bibfield  {title} {\bibinfo {title} {{Improved Predictions on Higgs-Starobinsky Inflation and Reheating with ACT DR6 and Primordial Gravitational Waves}},\ }\Eprint {https://arxiv.org/abs/2505.04615} {arXiv:2505.04615 [astro-ph.CO]} \BibitemShut {NoStop}%
\bibitem [{\citenamefont {Kallosh}\ \emph {et~al.}(2025)\citenamefont {Kallosh}, \citenamefont {Linde},\ and\ \citenamefont {Roest}}]{Kallosh:2025rni}%
  \BibitemOpen
  \bibfield  {author} {\bibinfo {author} {\bibfnamefont {R.}~\bibnamefont {Kallosh}}, \bibinfo {author} {\bibfnamefont {A.}~\bibnamefont {Linde}},\ and\ \bibinfo {author} {\bibfnamefont {D.}~\bibnamefont {Roest}},\ }\bibfield  {title} {\bibinfo {title} {{A simple scenario for the last ACT}},\ }\Eprint {https://arxiv.org/abs/2503.21030} {arXiv:2503.21030 [hep-th]} \BibitemShut {NoStop}%
\bibitem [{\citenamefont {Aoki}\ \emph {et~al.}(2025)\citenamefont {Aoki}, \citenamefont {Otsuka},\ and\ \citenamefont {Yanagita}}]{Aoki:2025wld}%
  \BibitemOpen
  \bibfield  {author} {\bibinfo {author} {\bibfnamefont {S.}~\bibnamefont {Aoki}}, \bibinfo {author} {\bibfnamefont {H.}~\bibnamefont {Otsuka}},\ and\ \bibinfo {author} {\bibfnamefont {R.}~\bibnamefont {Yanagita}},\ }\bibfield  {title} {\bibinfo {title} {{Higgs-Modular Inflation}},\ }\Eprint {https://arxiv.org/abs/2504.01622} {arXiv:2504.01622 [hep-ph]} \BibitemShut {NoStop}%
\bibitem [{\citenamefont {Gialamas}\ \emph {et~al.}(2025{\natexlab{a}})\citenamefont {Gialamas}, \citenamefont {Karam}, \citenamefont {Racioppi},\ and\ \citenamefont {Raidal}}]{Gialamas:2025kef}%
  \BibitemOpen
  \bibfield  {author} {\bibinfo {author} {\bibfnamefont {I.~D.}\ \bibnamefont {Gialamas}}, \bibinfo {author} {\bibfnamefont {A.}~\bibnamefont {Karam}}, \bibinfo {author} {\bibfnamefont {A.}~\bibnamefont {Racioppi}},\ and\ \bibinfo {author} {\bibfnamefont {M.}~\bibnamefont {Raidal}},\ }\bibfield  {title} {\bibinfo {title} {{Has ACT measured radiative corrections to the tree-level Higgs-like inflation?}},\ }\Eprint {https://arxiv.org/abs/2504.06002} {arXiv:2504.06002 [astro-ph.CO]} \BibitemShut {NoStop}%
\bibitem [{\citenamefont {Dioguardi}\ \emph {et~al.}(2025)\citenamefont {Dioguardi}, \citenamefont {Iovino},\ and\ \citenamefont {Racioppi}}]{Dioguardi:2025vci}%
  \BibitemOpen
  \bibfield  {author} {\bibinfo {author} {\bibfnamefont {C.}~\bibnamefont {Dioguardi}}, \bibinfo {author} {\bibfnamefont {A.~J.}\ \bibnamefont {Iovino}},\ and\ \bibinfo {author} {\bibfnamefont {A.}~\bibnamefont {Racioppi}},\ }\bibfield  {title} {\bibinfo {title} {{Fractional attractors in light of the latest ACT observations}},\ }\Eprint {https://arxiv.org/abs/2504.02809} {arXiv:2504.02809 [gr-qc]} \BibitemShut {NoStop}%
\bibitem [{\citenamefont {Salvio}(2025)}]{Salvio:2025izr}%
  \BibitemOpen
  \bibfield  {author} {\bibinfo {author} {\bibfnamefont {A.}~\bibnamefont {Salvio}},\ }\bibfield  {title} {\bibinfo {title} {{Independent connection in ACTion during inflation}},\ }\Eprint {https://arxiv.org/abs/2504.10488} {arXiv:2504.10488 [hep-ph]} \BibitemShut {NoStop}%
\bibitem [{\citenamefont {Antoniadis}\ \emph {et~al.}(2025)\citenamefont {Antoniadis}, \citenamefont {Ellis}, \citenamefont {Ke}, \citenamefont {Nanopoulos},\ and\ \citenamefont {Olive}}]{Antoniadis:2025pfa}%
  \BibitemOpen
  \bibfield  {author} {\bibinfo {author} {\bibfnamefont {I.}~\bibnamefont {Antoniadis}}, \bibinfo {author} {\bibfnamefont {J.}~\bibnamefont {Ellis}}, \bibinfo {author} {\bibfnamefont {W.}~\bibnamefont {Ke}}, \bibinfo {author} {\bibfnamefont {D.~V.}\ \bibnamefont {Nanopoulos}},\ and\ \bibinfo {author} {\bibfnamefont {K.~A.}\ \bibnamefont {Olive}},\ }\bibfield  {title} {\bibinfo {title} {{How Accidental was Inflation?}},\ }\Eprint {https://arxiv.org/abs/2504.12283} {arXiv:2504.12283 [hep-ph]} \BibitemShut {NoStop}%
\bibitem [{\citenamefont {He}\ \emph {et~al.}(2025)\citenamefont {He}, \citenamefont {Hong},\ and\ \citenamefont {Mukaida}}]{He:2025bli}%
  \BibitemOpen
  \bibfield  {author} {\bibinfo {author} {\bibfnamefont {M.}~\bibnamefont {He}}, \bibinfo {author} {\bibfnamefont {M.}~\bibnamefont {Hong}},\ and\ \bibinfo {author} {\bibfnamefont {K.}~\bibnamefont {Mukaida}},\ }\bibfield  {title} {\bibinfo {title} {{Increase of $n_s$ in regularized pole inflation \& Einstein-Cartan gravity}},\ }\Eprint {https://arxiv.org/abs/2504.16069} {arXiv:2504.16069 [astro-ph.CO]} \BibitemShut {NoStop}%
\bibitem [{\citenamefont {Kuralkar}\ \emph {et~al.}(2025)\citenamefont {Kuralkar}, \citenamefont {Panda},\ and\ \citenamefont {Vidyarthi}}]{Kuralkar:2025zxr}%
  \BibitemOpen
  \bibfield  {author} {\bibinfo {author} {\bibfnamefont {H.~J.}\ \bibnamefont {Kuralkar}}, \bibinfo {author} {\bibfnamefont {S.}~\bibnamefont {Panda}},\ and\ \bibinfo {author} {\bibfnamefont {A.}~\bibnamefont {Vidyarthi}},\ }\bibfield  {title} {\bibinfo {title} {{Effective Starobinsky pre-inflation}},\ }\Eprint {https://arxiv.org/abs/2504.15061} {arXiv:2504.15061 [gr-qc]} \BibitemShut {NoStop}%
\bibitem [{\citenamefont {Yogesh}\ \emph {et~al.}(2025)\citenamefont {Yogesh}, \citenamefont {Mohammadi}, \citenamefont {Wu},\ and\ \citenamefont {Zhu}}]{Yogesh:2025wak}%
  \BibitemOpen
  \bibfield  {author} {\bibinfo {author} {\bibnamefont {Yogesh}}, \bibinfo {author} {\bibfnamefont {A.}~\bibnamefont {Mohammadi}}, \bibinfo {author} {\bibfnamefont {Q.}~\bibnamefont {Wu}},\ and\ \bibinfo {author} {\bibfnamefont {T.}~\bibnamefont {Zhu}},\ }\bibfield  {title} {\bibinfo {title} {{Starobinsky like inflation and EGB Gravity in the light of ACT}},\ }\Eprint {https://arxiv.org/abs/2505.05363} {arXiv:2505.05363 [astro-ph.CO]} \BibitemShut {NoStop}%
\bibitem [{\citenamefont {Gialamas}\ \emph {et~al.}(2025{\natexlab{b}})\citenamefont {Gialamas}, \citenamefont {Katsoulas},\ and\ \citenamefont {Tamvakis}}]{Gialamas:2025ofz}%
  \BibitemOpen
  \bibfield  {author} {\bibinfo {author} {\bibfnamefont {I.~D.}\ \bibnamefont {Gialamas}}, \bibinfo {author} {\bibfnamefont {T.}~\bibnamefont {Katsoulas}},\ and\ \bibinfo {author} {\bibfnamefont {K.}~\bibnamefont {Tamvakis}},\ }\bibfield  {title} {\bibinfo {title} {{Keeping the relation between the Starobinsky model and no-scale supergravity ACTive}},\ }\Eprint {https://arxiv.org/abs/2505.03608} {arXiv:2505.03608 [gr-qc]} \BibitemShut {NoStop}%
\bibitem [{\citenamefont {Yin}(2025)}]{Yin:2025rrs}%
  \BibitemOpen
  \bibfield  {author} {\bibinfo {author} {\bibfnamefont {W.}~\bibnamefont {Yin}},\ }\bibfield  {title} {\bibinfo {title} {{Higgs-like inflation under ACTivated mass}},\ }\Eprint {https://arxiv.org/abs/2505.03004} {arXiv:2505.03004 [hep-ph]} \BibitemShut {NoStop}%
\end{thebibliography}%

\end{document}